\documentclass{article}

\usepackage{amsmath,amssymb}
\usepackage{graphicx}
\usepackage{subcaption}
\usepackage{hyperref}
\usepackage[margin=1in]{geometry}
\usepackage[toc,page]{appendix}
\usepackage{authblk}

\title{Symmetry-based perturbation theory for electronic structure
  calculations}

\author[1]{Hiromichi Nishimura\thanks{hiromichi.nishimura@boeing.com}}
\affil[1]{Applied Mathematics, Boeing Technology Innovation,
  Huntington Beach, California 92647, USA}

\author[1]{Nam Nguyen} 

\author[2]{Tanvi Gujarati} \affil[2]{IBM Quantum, IBM Almaden
  Research Center, San Jose, California 95120, USA}

\author[3]{Mario Motta} \affil[3]{IBM Quantum, IBM T.~J.~Watson
  Research Center, Yorktown Heights, New York 10598, USA}

\begin{document}

\maketitle

\begin{abstract}
  We develop a multireference perturbation theory for electronic
  structure calculations based on symmetries of the Hamiltonian. The
  reference Hamiltonian in the symmetry-based perturbation theory
  (SBPT) is chosen such that it possesses more symmetries than the
  original Hamiltonian, leading to a larger reduction of computational
  resources in terms of both the number of configurations in the
  configuration-interaction expansion and the number of required
  qubits in quantum computing applications. We provide approximate,
  scalable solutions for the second-order correction, as well as an
  application to selected configuration interaction. We show that SBPT
  is an extension of other existing multireference perturbation
  theories and that it can give better results for some molecular
  systems in a robust way.
\end{abstract}

\section{Introduction}
\label{sec:introduction}

Computing the electronic structure of molecular systems is of great
importance in both fundamental science and industry. It can explain
the behavior and properties of molecules and can guide the search for
new materials. Designs for corrosion-resistant materials, drugs, and
batteries, for example, can be aided by computational chemistry to
accelerate the search for better products through experiments
\cite{Williams2015MgCorr,
  Ebenso2021Corr,Sai2014photo-oxi,Zhou2026DrugDisc,
  Sadybekov2023DrugDisc,Bhatt2015Battery}.

{\it{Ab initio}} calculations to solve the electronic structure
problem use the Schr\"{o}dinger equation for $N$-electron systems with
fixed nuclei. Exact solutions for molecular systems do not exist,
except in special circumstances like noninteracting systems and the
Bethe ansatz for one-dimensional systems \cite{Bethe1931}. Even exact
numerical solutions by the full configuration-interaction (FCI) method
are approximated, as they are only exact within a finite basis
set. The exact solution is, however, not necessary or even relevant in
many cases; achieving sufficient accuracy for the properties of
interest is what matters. One of the goals of quantum chemistry is to
model or approximate a molecular system such that we can capture the
correct physics and obtain a result with desired accuracy.

Perturbation theory (PT) is one of many approximation methods with
substantially low computational cost. It first solves a reference
Hamiltonian by ignoring part of the Hamiltonian, and then it
incorporates perturbative corrections using expansions such as
asymptotic series. There are now many perturbation methods that differ
by the reference Hamiltonian, but they can be categorized as either
single-reference PT (SRPT) or multireference PT (MRPT). In a SRPT,
such as Moller-Plesset PT \cite{Mller1934NoteOA} and Epstein-Nesbet PT
\cite{Epstein1926TheSE,Nesbet1955ConfigurationII}, a solution to the
reference Hamiltonian is a single electron configuration.  A single
mean-field wave function such as the Hartree-Fock (HF) state is a good
approximation to a solution for single-reference systems, and a SRPT
typically gives accurate results with very low cost.

Many interesting problems, however, require a multireference
description and thus invalidate single-reference approaches. They
include bond breaking, transition-metal complexes, magnetic systems,
and excited states \cite{Bulik2015CCSD, Sharma2014FeSclusters,
  Kurashige2013Mn4CaO5, Zhendong2019Pcluster,
  Norman2016Herbertsmithite, Cohen2008DFT, Anisimov1997LDAU} .  MRPTs
aim to overcome the issue by increasing the size of the reference
Hamiltonian.  Solutions to the reference Hamiltonian in standard MRPTs
are typically full solutions in a smaller orbital basis called
complete active space (CAS).  A standard MRPT theory based on CAS,
such as CAS perturbation theory (CASPT) \cite{Roos1982ASM} and
$n$-electron valence state perturbation theory (NEVPT)
\cite{Angeli2001IntroductionON,Angeli2001NelectronVS,
  Angeli2002nelectronVS}, is designed such that the reference
Hamiltonians can capture strong (static) correlations, while
perturbative corrections incorporate weak (dynamical)
correlations. The partition of the Hamiltonian aims to encapsulate
correct physics in the reference Hamiltonian while obtaining accurate
results through a perturbative expansion.  The method can be
systematically improved as we increase the size of the CAS in exchange
for a larger computational cost, and it becomes exact as CAS
approaches FCI.

Quantum computing is yet another approach that could obtain
high-fidelity results with low computational cost. For example,
quantum phase estimation (QPE) \cite{Kitaev1995QuantumMA} could
provide a result comparable to FCI with only polynomial time
complexity on a noiseless quantum computer, thus possibly achieving a
computational speedup \cite{AspuruGuzik2005SimulatedQC}.
Fault-tolerant implementation to achieve optimal complexity in the
presence of errors relies heavily on efficient algorithms for quantum
time evolution and state preparation, and it is an active research
area. As a heuristic method, the quantum computing application to the
electronic structure problems is possible without error correction for
noisy intermediate-scale quantum (NISQ) devices. The NISQ algorithms
aim to prepare states of interest with a quantum circuit depth that is
orders of magnitude shallower than QPE by introducing variational
parameters in the variational quantum eigensolver
\cite{Peruzzo2014VQE} or by employing approximate circuits such as the
local unitary cluster Jastrow ansatz \cite{motta2023bridging}. There
are now many NISQ algorithms for electronic structure problems, and
some well-studied ones include the Krylov method
\cite{yu2025quantumcentric, Nobuyuki2025KrylovDiag}, quantum selected
configuration interaction (QSCI) \cite{dmn4-snfx,
  doi:10.1021/acs.jctc.4c00846, Kenji2025Hamiltonian,
  75pv-hbrx, shirai2025enhancing}, and sample-based quantum
diagonalization (SQD) \cite{robledo2024chemistry,
  kaliakin2024accurate, liepuoniute2024quantum, smith2025gef,
  barroca2025scaling, Barison_2025, shivpuje2025sample}.  As another
heuristic method, the implementation of NEVPT in quantum computing
applications has been studied in
Refs.~\cite{doi:10.1021/acs.jpca.2c07653, Krompiec2022Strongly}.

This paper can be seen as another quantum computing application of
perturbation theory in quantum chemistry. But there is more. We
introduce a MRPT by extending the existing MRPT framework. Our
proposed reference Hamiltonian is constructed from symmetry
considerations, and we call this symmetry-based perturbation theory
(SBPT). We review the symmetry of the electronic Hamiltonian in
Sec.~\ref{sec:electronic_hamiltonian} and show how we can exploit the
symmetry to construct the reference Hamiltonian in
Sec.~\ref{sec:sbpt}. Our approach can be used as a classical or
quantum computational method.

$\mathbb{Z}_2$ symmetry is particularly important because of the
adaptation of qubit tapering, as discussed in
Sec.~\ref{sec:quantum_computing}.  We also discuss the implementation
of selected configuration interaction (SCI) to circumvent the
shortcomings of perturbation theory in Sec.~\ref{sec:sci_sbpt}.  We
demonstrate our methods on some small molecular systems in
Sec.~\ref{sec:examples}.  These results can be obtained exactly
through quantum computations in a noiseless simulation with
significant resource reductions. Our initial study focuses on simple
molecules to demonstrate the outperformance of SBPT over other MRPT
methods in a concrete way. We then conclude in
Sec.~\ref{sec:conclusions}.

\section{Symmetries of the electronic Hamiltonian}
\label{sec:electronic_hamiltonian}

In this section, we establish our notations and review the aspects of
symmetry in quantum chemistry needed for this paper. For more details,
readers can refer to
Refs.~\cite{Szab1982ModernQC,Helgaker2000MolecularET} for molecular
orbital theory and
Refs.~\cite{Cotton1971ChemicalAO,Tinkham1964GroupTA} for point group
symmetries.

We consider the electronic structure of a molecule, where the nuclei
are treated as stationary using the Born-Oppenheimer
approximation. The Hamiltonian, called the electronic Hamiltonian, in
atomic units is then given by
\begin{equation}
  H = - \sum_i \frac{1}{2} \nabla^2_i - \sum_{i, l} \frac{Z_l}{\left|
      \mathbf{r}_i - \mathbf{R}_l \right|} + \sum_{i > j}
  \frac{1}{\left| \mathbf{r}_i -\mathbf{r}_j \right|}
  \label{electronic_hamiltonian}
\end{equation}
where $\mathbf{R}_l$ and $Z_l$ are the position and the atomic number
of $l$-th nucleus, and $\mathbf{r}_i$ is the position of the $i$-th
electron. One of the goals in quantum chemistry is to solve the
eigensystem, namely for the $n$-the eigenvalue of the Hamiltonian:
\begin{equation}
  H \left| \Psi_n\right> = E_n \left| \Psi_n \right>
  \label{eigenvalue}
\end{equation}
where the eigenstate $\left| \Psi_n \right>$ is an $N$-electron
wavefunction, which is a function of the spatial and spin degrees of
freedoms for $N$ electrons.

{\bf{Symmetries -- }} The eigenstate exhibits two different kinds of
symmetries. First it is antisymmetric with respect to the exchange of
two electrons due to the spin-statistics theorem for fermions. This
property makes the eigenvalue problem notoriously difficult: for
example, it gives rise to the sign problem in quantum Monte Carlo
simulations.  Second it respects the symmetries of the electronic
Hamiltonian.  We consider three symmetries in the electronic
Hamiltonian: $U(1)$ charge symmetry, $SU(2)$ spin symmetry, and a
point group symmetry. We will discuss each symmetry in a little more
detail below.  The unitary operator $U_a$, labeled by $a$ and
belonging to the symmetry group, commutes with the Hamiltonian,
$\left[ H, U_a \right] = 0$. Therefore the eigenstate is an
irreducible representation $\theta$ of the symmetry group, and we now
denote the $n$-th eigenstate in the irrep $\theta$ as
$\left|\Psi_{n_\theta} \right>$. We use the terms irreducible
representation and irrep interchangeably. This reducibility can
restrict the form of the $N$-electron wavefunction and simplify the
problem.

\begin{figure}
  \centering
  \includegraphics[width=7cm]{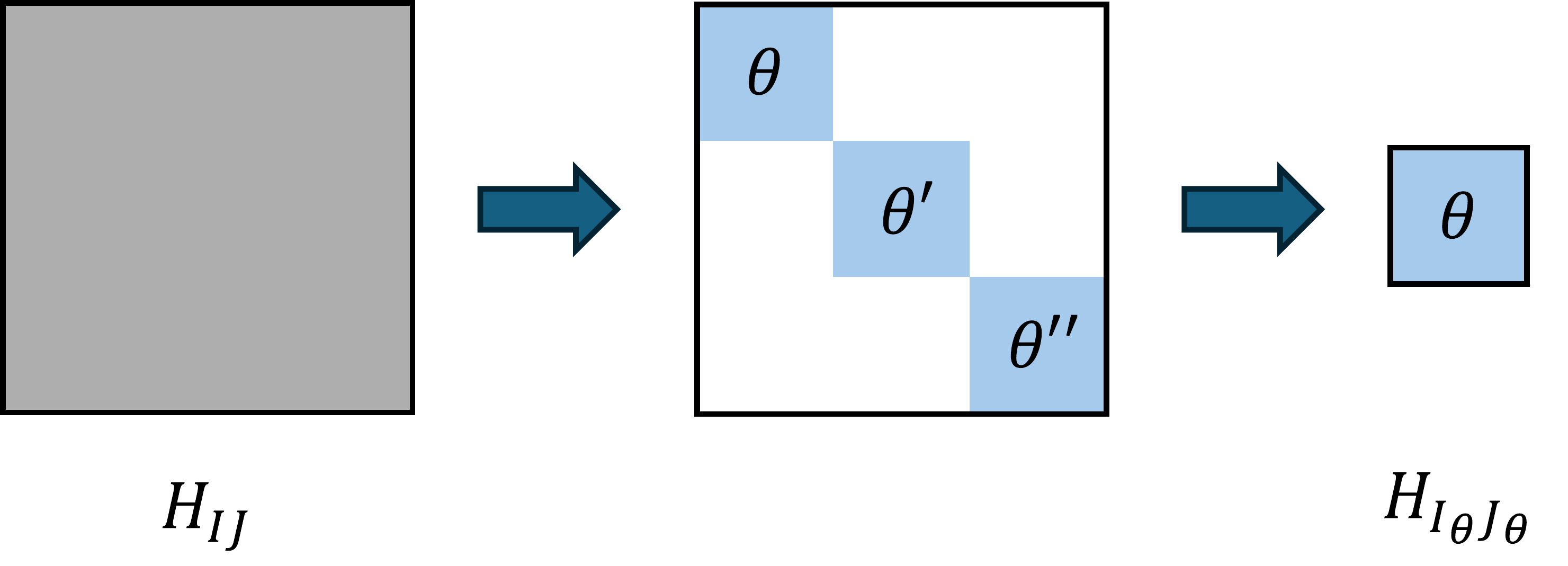}
  \caption{The matrix on the left is a $D \times D$ Hamiltonian
    matrix, $H_{IJ} = \left< \Phi_I \right| H \left| \Phi_J
    \right>$. The Hamiltonian can be reduced to a block-diagonal form
    using the symmetries of $H$, where unshaded elements are all zero.
    Each block matrix is labeled by its irreducible representation
    $\theta$ and given by a $D_{\theta} \times D_{\theta}$ matrix,
    $H_{I_{\theta}J_{\theta}} = \left<\Phi_{I_{\theta}} \right| H
    \left| \Phi_{J_{\theta}} \right>$.}
  \label{fig:reduction}
\end{figure}

To illustrate how we typically take advantage of the symmetry, we
consider a set of $N$-electron wavefunctions $\left| \Phi_I \right>$
with $I=1, 2, \dots, D$ as basis functions. The finite basis
approximation is necessary except in special exactly solvable cases,
such as the hydrogen atom. To obtain the eigenvalues, we simply
diagonalize the $D \times D$ Hamiltonian matrix,
$H_{IJ} = \left< \Phi_I \right| H \left| \Phi_J \right>$. On the other
hand, if we can choose a finite set of basis functions that capture
the symmetry of the Hamiltonian, then we can completely reduce it to a
block-diagonal form, where the basis functions are now in an
irreducible representation, $\left| \Phi_{I_{\theta}} \right>$ with
$I_{\theta} = 1, 2, \dots, D_{\theta}$.  We use the same notation
$\Phi$ for brevity, but the new basis
$\left| \Phi_{I_{\theta}} \right>$ can be a linear combination of the
original basis $\left| \Phi_{I} \right>$: one such example is the
configuration state function basis mentioned below.  In this new
basis, we only need to diagonalize the subspace
$H_{I_{\theta} J_{\theta}} = \left< \Phi_{I_{\theta}} \right| H \left|
  \Phi_{J_{\theta}} \right>$ to obtain the $n_{\theta}$-th eigenstate
$\left|\Psi_{n_\theta} \right>$:
\begin{equation}
  \left|\Psi_{n_\theta} \right> = \sum_{I_\theta} c_{I_{\theta}}
  \left| \Phi_{I_{\theta}} \right>.
  \label{eigenstate}
\end{equation}
This reduction process is illustrated in Fig.~\ref{fig:reduction}. The
symmetry of the electronic Hamiltonian can therefore reduce the
computational cost of solving the eigensystems, if we can construct
the basis functions in the irreducible representations
$\left| \Phi_{I_{\theta}} \right>$. Next, we discuss how to construct
such bases.

{\bf{Basis states -- }} One way to construct a basis set of
$N$-electron wavefunctions in general is to start from a set of
one-electron wavefunctions. A one-electron wavefunction is a product
of a spatial orbital function $\phi_p$ and a spin function: together
it is called a spin orbital. We denote the spin orbital as $\psi_p$.
An $N$-electron wavefunction can be then constructed by the
antisymmetrization of a tensor product of $N$ spin orbitals, called a
Slater determinant or a configuration.

In this paper, we only consider an abelian subgroup $S$ of the
symmetry in the electronic Hamiltonian. The reason is that we can then
construct every Slater determinant in an irreducible representation of
the abelian subgroup.  The $SU(2)$ spin symmetry, for example, is
non-abelian. For the restricted determinant where the spatial orbitals
are the same for spin up and down, one can construct a spin eigenstate
using a symmetry-adapted linear combination (SALC) of Slater
determinants. This SALC is called a spin-adapted configuration or a
configuration state function (CSF).  In general, irreducible
representations of a non-abelian group cannot be written as a single
Slater determinant and require a SALC of multiple Slater
determinants. While the use of SALC reduces the dimension of the
Hamiltonian matrix, it may not be easy or computationally efficient to
construct such a basis. There is, however, a recent development in
quantum computing application to prepare ground states efficiently
using CSFs in the quantum subspace expansion
\cite{doi:10.1021/acs.jctc.6c00017}.

The $U(1)$ charge symmetry, which is abelian, gives rise to the
conservation of the total number $N$ of electrons.  It is clear that
the Slater determinant is already an eigenstate of the number operator
$\hat{N}$. In general, when the number of electrons $N_a$ in a set
$L_a$ of spin orbitals is conserved, such as spin up ($\uparrow$) or
spin down ($\downarrow$) due to the $U(1)$ subgroup of SU(2) spin
symmetry, the corresponding unitary symmetry operation is a $U(1)$
rotation. To compute a spin singlet eigenstate, for example, we only
need to consider the Slater determinants
$\left| \Phi_{I_\theta} \right>$ with the eigenvalues
$N_{\uparrow} = N_{\downarrow} = N/2$. The number of relevant
configurations with $M$ orbitals grows at worst, when $N=M$, as
$\sim 2^{2N}/N$ rather than $\sim 2^{2N}$, due to $U(1)$ symmetries,
which helps reduce the computational cost.

The other symmetry, point group symmetry, is present when a molecule
has a symmetric spatial structure, which is invariant under coordinate
transformations such as rotations or inversions. To construct spatial
orbitals, we typically employ a set of atomic orbitals and use a
linear combination of them. We then minimize the energy of a Slater
determinant as a function of its occupied orbitals self-consistently
using the Hartree-Fock (HF) equation. Since the HF equation respects
the point-group symmetry of the full Hamiltonian, the converged
solutions should be in the irreducible representation of the point
group.  We can enforce this condition by using the projection operator
to construct SALCs of atomic orbitals. Again we only consider the
unitary representation of the abelian subgroup of the point group
symmetry.

An irreducible representation is defined by its characters on the
conjugacy classes of the group, i.e., corresponding to a row in the
character table for the point group symmetry. For abelian groups,
every irrep is one dimensional and is uniquely determined by its
characters, or equivalently eigenvalues, of the group generators.
Acting a unitary operator $U_a$ in the abelian group $S$ on a spin
orbital $\psi_p$, we have
$U_a \left| \psi_p \right> = e^{i \theta_p(a)} \left| \psi_p \right>$,
where $e^{i\theta_p}$ is the character. For the $\mathbb{Z}_2$
symmetry, $e^{i \theta_p(a)} = 1$ or $-1$. Acting the operator on the
$N$-electron Slater determinant, we obtain
\begin{equation}
  U_a \left| \Phi_{I_{\theta}} \right> = e^{i
    \theta(a)}  \left| \Phi_{I_{\theta}} \right> = \prod_{p \in L_a}
  e^{i \theta_p (a) \hat{N}_p} \left| \Phi_{I_{\theta}} \right>
\end{equation}
where $\hat{N}_p$ is the number operator for the spin orbital
$\psi_p$, and $L_a$ is a set of orbitals that have a non-trivial
character, i.e., $\theta_p(a) \neq 0 $ for $p \in L_a$.  For the
Slater determinant $\left| \Phi_{I_{\theta}} \right>$ in an irrep
$\theta$, the number of electrons occupying in the set $L_a$ is
constrained by the equation
\begin{equation}
  \theta (a) = \sum_{p \in L_a} \theta_p (a) \left<
    \Phi_{I_{\theta}} \right| \hat{N}_{p} \left| \Phi_{I_{\theta}}
  \right>
\end{equation}
for each operator $U_a$. Again the equation is completely determined
by the characters of the group generators.  We can therefore construct
Slater determinants as the $N$-electron basis functions
$\left| \Phi_{I_{\theta}} \right>$ in the one-dimensional irrep of the
abelian subgroup of the symmetries in the electronic Hamiltonian.

{\bf{Second quantization -- }} The matrix $H_{IJ}$ is sparse because
the electronic Hamiltonian Eq.~\eqref{electronic_hamiltonian} is a
two-body Hamiltonian.  For example, the coupling between two
configurations with more than two different spin orbitals is zero by
the Slater-Condon rules.  For this reason, instead of explicitly
constructing the full matrix $H_{IJ}$ as in Fig.~\ref{fig:reduction},
we use the second-quantized Hamiltonian
\begin{equation}
  H = \sum_{pq} h_{pq} a^{\dagger}_p a_q + \frac{1}{2} \sum_{pqrs}
  h_{pqrs} a^{\dagger}_p a^{\dagger}_q a_r a_s.
  \label{2nd_q}
\end{equation}
This is an efficient way to store the same information without
computing all nonzero elements of the Hamiltonian matrix. The
coefficients $h_{pq}$ and $h_{pqrs}$ are the one-body and two-body
integrals:
\begin{equation}
  \begin{aligned}
    h_{pq} &= \left< \psi_p \right| \left( - \frac{\nabla^2}{2} -
             \sum_{l} \frac{Z_l}{\left| \mathbf{r} - \mathbf{R}_l
             \right|} \right) \left| \psi_q \right> \\
    h_{pqrs} &= \left<\psi_p \psi_q \right| \frac{1}{\left| \mathbf{r}
               - \mathbf{r}' \right|} \left| \psi_s \psi_r \right>
  \end{aligned}
\end{equation}
where $\left| \mathbf{r} - \mathbf{r}' \right|$ is the distance
between the two electrons and the operator $a^\dagger_p$ ($a_q$)
creates (destroys) an electron in the spin-orbital $p$ ($q$).

The unitary operator $U_a$ commutes not only with the Hamiltonian,
$[H, U_a] = 0$, but also with each fermionic operator,
$\mathcal{O} = a^{\dagger}_p a_q$ or
$a^{\dagger}_p a^{\dagger}_q a_r a_s$, in the second-quantized
Hamiltonian,
\begin{equation}
  \left[\mathcal{O}, U_a \right] = 0.
  \label{fop_commutation}
\end{equation}
The reason is the following. For any basis function
$\left| \Phi_{I_{\theta}} \right>$ in the irrep $\theta$, the
operation of $a^{\dagger}_p a_q$ or
$a^{\dagger}_p a^{\dagger}_q a_r a_s$ transforms it to another basis
function in the same irrep $\theta$, because $\theta_p = \theta_q$ or
$\theta_p + \theta_q = \theta_r + \theta_s$, respectively, from the
definition of one-body and two-body integrals above.  Any algorithm
that uses fermionic operators in the Hamiltonian, such as the FCI
calculation of the ground state energy with Davidson's method or what
we discuss in this paper, ensures computation in the subspace of the
irreducible representation, and this is again the consequence of
symmetry reduction.

\section{Symmetry-based perturbation theory}
\label{sec:sbpt}

As we have discussed in the previous section, the symmetries of the
electronic Hamiltonian can be used to reduce computational cost.  We
now propose a perturbative method, where the choice of the reference
Hamiltonian is guided by the symmetry.  The idea of symmetry-based
perturbation theory (SBPT) is simple: we split the second quantized
Hamiltonian in Eq.~\eqref{2nd_q} into two parts,
\begin{equation}
  H =   H_{\rm{ref}} + H_{\rm{pert}}
  \label{h_splitting}
\end{equation}
such that the reference Hamiltonian $H_{\rm{ref}}$ has a larger
abelian group $S_{\rm{ref}}$ than the original abelian group $S$.

In this section, we develop this idea. We first consider the types of
approximate symmetries relevant for SBPT in
Sec.~\ref{sec:approximate_symmetries}.  We then show how we split the
Hamiltonian based on the approximate symmetry in
Sec.~\ref{sec:optimal_partitioning} and compute the second-order
correction in Sec.~\ref{sec:second_order}. Lastly we discuss the
relation to other multireference perturbation theories in
Sec.~\ref{sec:other_pt} and possible issues in
Sec.~\ref{sec:possible_issues}.

\subsection{Approximate symmetries}
\label{sec:approximate_symmetries}

\begin{figure}
  \centering
  \includegraphics[width=4cm]{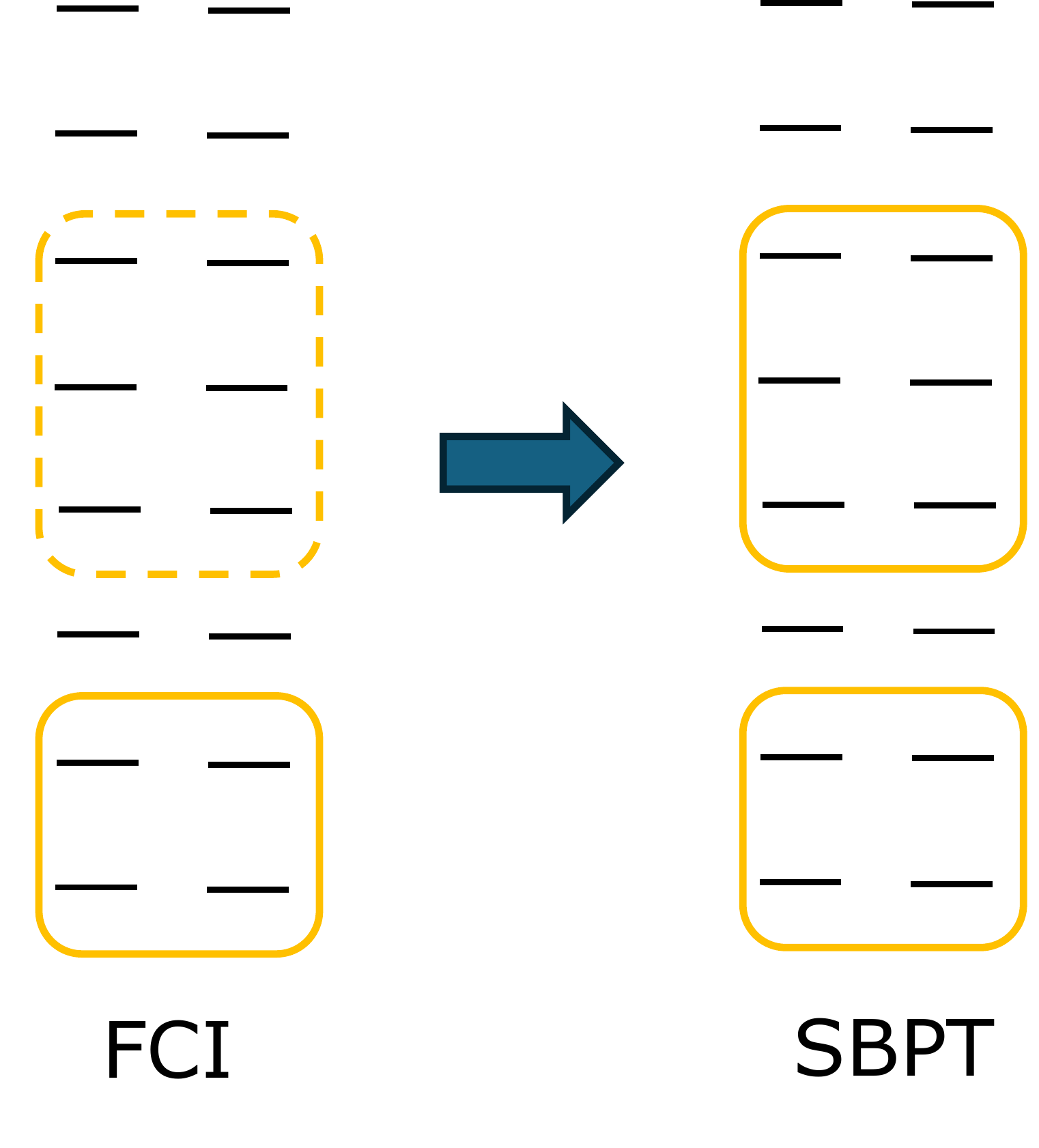}
  \caption{A schematic orbital energy diagram, with each horizontal
    black line representing a spin orbital $\psi_p$. The orbitals in
    the solid box transform nontrivially under a symmetry operation in
    the abelian group.  The number of such operations and the number
    of solid boxes in SBPT are greater than in FCI---the more the
    number of boxes, the smaller the computational cost to solve for
    the reference Hamiltonian.}
  \label{fig:sbpt}
\end{figure}

We consider an orbital energy diagram shown in Fig.~\ref{fig:sbpt},
where the horizontal black line indicates a spin orbital $\psi_p$.
The solid box indicates a set $L_a$ of spin orbitals, where the number
of electrons $N_a$ occupying the spin orbitals in the full
configuration interaction (FCI) is constrained due to the exact
abelian symmetry $S$, reducing the number of possible Slater
determinants as discussed in the previous section.

Suppose that there is another set $L_b$ of orbitals that are
nontrivially coupled, as indicated by a dashed box.  A nontrivial
coupling means that the number of electrons $N_{b}$ occupying the set
$L_b$ is approximately constrained in the FCI. There are at least two
cases in which the orbitals are nontrivially coupled: (I) The orbitals
transform nontrivially under an approximate point-group symmetry. (II)
The orbitals couple weakly with the rest of the orbitals.

The first case (I) may occur when the geometry is slightly deviated
from the symmetric structure. We study such a case in
Sec.~\ref{sec:h2o}. It is then natural to construct a reference
Hamiltonian that makes the approximate symmetry exact. We note that
this type of approximate symmetry is related to quasi-symmetry in
condensed matter physics
\cite{Guo2021QuasisymmetryPT,PhysRevB.107.125145,PhysRevLett.133.026402}.
For the second case (II), we point out that the total number of
electrons $N_b$ is conserved if the spin orbitals $\psi_p$ with
$p \in L_b$ do not couple at all with the rest of the orbitals,
$\psi_q$ with $q \not\in L_b$. By treating a small interaction between
the $\psi_{p}$ and the rest of the orbitals as a perturbation, the
total number of electrons in $L_b$ is conserved in the reference
Hamiltonian, which then possesses a new symmetry for the particle
number.

In this regard the two cases are the same from the symmetry
perspective. The first case (I) promotes the approximate point-group
symmetry to the exact one, while the second case (II) creates a new
particle number symmetry in the reference Hamiltonian. In either case,
we make the approximate symmetry as an exact symmetry and construct
the reference Hamiltonian that respects the new augmented symmetry
$S_{\rm{ref}}$, as illustrated in the orbital energy diagram with a
new solid box on the right in Fig.~\ref{fig:sbpt}. 

\subsection{Optimal partitioning}
\label{sec:optimal_partitioning}

Given a new augmented symmetry $S_{\rm{ref}}$, the idea of SBPT is to
split the Hamiltonian as in Eq.~\eqref{h_splitting}, where every
fermionic operator $\mathcal{O}_{\rm{ref}}$ in $H_{\rm{ref}}$ has the
commutation relationship
$\left[\mathcal{O}_{\rm{ref}}, U_a \right] = 0$ for any symmetry
operator $U_a$ in $S_{\rm{ref}}$.  It is easy to see that the
reference Hamiltonian in this construction is not unique.

We can further require that each fermionic operator
$\mathcal{O}_{\rm{pert}}$ in $H_{\rm{pert}}$ does not commute with at
least one of the new symmetry operators.  That is, we require that
each fermionic operator in $H_{\rm{pert}}$ satisfies
\begin{equation}
  \left[\mathcal{O}_{\rm{pert}}, U_a \right] \neq 0
  \label{optimal_partitioning}
\end{equation}
for some $U_a$ in $S_{\rm{ref}}$.  This additional requirement forbids
adding a fermionic operator that respects $S_{\rm{ref}}$ into
$H_{\rm{pert}}$, or in other words it ensures each operator in
$H_{\rm{pert}}$ breaks $S_{\rm{ref}}$.  This means that each fermionic
term in $H_{\rm{pert}}$ transforms a basis function
$\left| \Phi_{I_\theta} \right>$ in the irrep $\theta$ into another
basis function $\left| \Phi_{I_{\theta'}} \right>$ with a different
irrep $\theta' \neq \theta$:
\begin{equation}
  \mathcal{O}_{\rm{pert}} \left| \Phi_{I_\theta} \right> = \gamma \left|
    \Phi_{I_{\theta'}} \right>
\end{equation}
where $\theta' = \theta - \theta_q + \theta_p$ for
$\mathcal{O}_{\rm{pert}} = a^{\dagger}_p a_q$ and
$\theta' = \theta - \theta_s - \theta_r + \theta_q + \theta_p$ for
$\mathcal{O}_{\rm{pert}} = a^{\dagger}_p a^{\dagger}_q a_r a_s$, and
$\gamma = \pm 1$ or $0$.

The extra requirement in Eq.~\eqref{optimal_partitioning} is not
necessary to perform the perturbative expansion, but it gives rise to
a few desirable properties. First, for a given symmetry
$S_{\rm{ref}}$, it is clear that this is the optimal way to split the
Hamiltonian, where the reference Hamiltonian uniquely gives the
largest norm $|| H_{\rm{ref}} ||$. We therefore call the way of
splitting in Eq.~\eqref{optimal_partitioning} the optimal
partitioning. Having a small norm for the perturbation is a necessary
condition for a meaningful perturbative expansion at lower order.
Other nice features include that the first-order correction is always
zero, and we can make an efficient approximation for the second-order
correction, as we show in the next subsection.

In order to compute the energy spectrum in an irreducible
representation $\theta$, it is convenient to write a Hamiltonian as
$H = \sum_{\theta'} V_{\theta', \theta}$, where the fermionic operator
in $V_{\theta', \theta}$ transforms a state of our interest in the
representation $\theta$ into a state with $\theta'$.  In the optimal
partitioning, we have $H_{\rm{ref}} = V_{\theta,\theta}$, and
\begin{equation}
  H_{\rm{pert}} = \sum_{\theta' \neq \theta} V_{\theta', \theta}
  \label{v}
\end{equation}
where the summation is over $\theta'$ only and $\theta' = \theta$ is
excluded.  The number of fermionic operators in $V_{\theta', \theta}$
is at most the number of fermionic operators in $H_{\rm{pert}}$, which
grows only polynomially with the size of the problem. This fact is
important for the higher-order corrections.

\subsection{Perturbative expansion up to second order}
\label{sec:second_order}

The optimal partitioning in Eq.~\eqref{optimal_partitioning} defines a
unique reference Hamiltonian for a given symmetry $S_{\rm{ref}}$, and
we can perform the standard time-independent Rayleigh-Schr\"{o}dinger
perturbation theory (PT).  We consider an $n$-th state in an
irreducible representation $\theta$ of the full Hamiltonian, denoted
as $\left| \Psi_{n_{\theta}} \right>$ with the energy
$E_{n_{\theta}}$, and perform the perturbative expansion up to the
second-order correction using the optimal partitioning.

The leading-order contribution of $E_{n_{\theta}}$ is an $n$-th
eigenvalue of the reference Hamiltonian in the irreducible
representation $\theta$:
\begin{equation}
  H_{\rm{ref}} \left| \Psi^{(0)}_{n_{\theta}} \right> =
  E^{(0)}_{n_{\theta}} \left|
    \Psi^{(0)}_{n_{\theta}} \right>
  \label{e0}
\end{equation}
where the superscript denotes the order of perturbative expansion. The
augmented symmetry in $H_{\rm{ref}}$ can reduce the computational cost
for this eigenvalue problem.

The first-order correction is given as the expectation value of
$H_{\rm{pert}}$ with respect to the unperturbed state. Acting the $V$
operator in Eq.~\eqref{v} on the unperturbed state, we obtain an
unnormalized state in an irreducible representation
$\theta' \neq \theta$:
\begin{equation}
  V_{\theta', \theta} \left| \Psi^{(0)}_{n_{\theta}} \right> = \left|
    \Xi_{n_{\theta'}} \right>.
  \label{xi}
\end{equation}
Using this, we can show that (1) the leading-order contribution for
the ground-state energy gives an upper bound, $E_0 \leq E^{(0)}_0$,
due to the variational principle,
$E_0 \leq \left<\Psi^{(0)}_{0} \right| H \left| \Psi^{(0)}_{0} \right>
= \left< \Psi^{(0)}_{0} \right| H_{\rm{ref}} \left| \Psi^{(0)}_{0}
\right> = E^{(0)}_0$, and (2) the first-order correction is always
zero,
\begin{equation}
  E^{(1)}_{n_{\theta}} = \left< \Psi^{(0)}_{n_{\theta}} \right|
  H_{\rm{pert}} \left| \Psi^{(0)}_{n_{\theta}} \right> = \sum_{\theta'
    \neq \theta}\left. \left< \Psi^{(0)}_{n_{\theta}} \right|
    \Xi_{n_{\theta'}} \right> = 0,
  \label{1st_order}
\end{equation}
because the states in different irreducible representations are
orthogonal.  The optimal partitioning is optimal not only because the
norm of $H_{\rm{pert}}$ is the smallest for a given symmetry, but also
because the first-order correction is already incorporated into the
leading contribution, potentially reducing computation cost.

In the optimal partitioning, nontrivial contributions start from the
second-order correction:
\begin{equation}
  \begin{aligned}
    E^{(2)}_{n_{\theta}}
    =
      \sum_{\theta' \neq \theta} \sum_{m_{\theta'}} \frac{\left|
    \left< \Psi^{(0)}_{m_{\theta'}} \right| H_{\rm{pert}}
      \left| \Psi^{(0)}_{n_{\theta}} \right>
      \right|^2}{E^{(0)}_{n_{\theta}} - E^{(0)}_{m_{\theta'}}}
    =
      \sum_{\theta' \neq \theta} \sum_{m_{\theta'}} \frac{\left|
    \left. \left< \Psi^{(0)}_{m_{\theta'}} \right|
      \Xi_{n_{\theta'}} \right> \right|^2}{E^{(0)}_{n_{\theta}} -
    E^{(0)}_{m_{\theta'}}}
    \label{2nd_order}
  \end{aligned}
\end{equation}
where we have used Eqs.\eqref{v}, \eqref{xi}, and \eqref{1st_order} in
the second equality.  This is a familiar expression for the
second-order correction, only rewritten to show indices for
irreducible representations.

The double summations indicate that we need all $m$-th excited states
of $H_{\rm{ref}}$ in each irreducible representation
$\theta' \neq \theta$.  If the dimensions of irreducible
representations are small, then it is possible to diagonalize each
subspace to obtain the unperturbed excited states
$\left| \Psi^{(0)}_{m_{\theta'}} \right>$.  Moller-Plesset PT is such
an example, where the dimension is $1$, and the excited states are
simply all possible Slater determinants.  For multireference problems,
the irreducible representation with a small dimension may fail to
capture important configurations at the leading order, and the
higher-order corrections may not be able to remedy this lack of
multireference character.  In these cases, it is crucial to contain
many relevant configurations at leading order to capture the correct
physics.  Therefore the dimension of irreducible representations of
$H_{\rm{ref}}$ needs to be sufficiently large, and as a result,
obtaining all excited states becomes computationally demanding for
multireference problems. Therefore we need to make an approximation
for the second-order correction, just as in other multireference PTs.

There are several ways to approximate the second-order correction.
All approaches use an approximation for the excited states to simplify
the calculation. One approach is to use a mean-field approximation
where we replace each excited state with a single Slater determinant:
\begin{equation}
  E^{(2):\rm{EN}}_{n_{\theta}} = \sum_{\theta' \neq \theta}
  \sum_{I_{\theta'}} \frac{\left| \left< \Phi_{I_{\theta'}} \right|
      H_{\rm{pert}} \left| \Psi^{(0)}_{n_{\theta}}
      \right>\right|^2}{E^{(0)}_{n_{\theta}} - E_{I_{\theta'}}}
  \label{en}
\end{equation}
where $\left| \Phi_{I_{\theta'}} \right>$ is the Slater determinant
and
$E_{I_{\theta'}} = \left< \Phi_{I_{\theta'}} \right| H \left|
  \Phi_{I_{\theta'}} \right>$. This approximation is a generalization
of the multireference Epstein-Nesbet PT as we discuss in
Sec.~\ref{sec:other_pt}, so we call this second-order correction the
Epstein-Nesbet (EN) approximation.  We could also improve this
mean-field approximation by using excited states of another reference
Hamiltonian with a small active space.

Another approach employs a contraction method, where we approximate
the density matrix in Eq.\eqref{2nd_order} with a single state:
\begin{equation}
  \sum_{m_{\theta'}} \left| \Psi^{(0)}_{m_{\theta'}} \right> \left<
    \Psi^{(0)}_{m_{\theta'}} \right|
  \simeq
  \frac{\left| \Xi_{n_{\theta'}} \right> \left< \Xi_{n_{\theta'}}
    \right|}{\left. \left< \Xi_{n_{\theta'}} \right| \Xi_{n_{\theta'}}
    \right>}.
\end{equation}
This is analogous to the strongly contracted (SC) method in NEVPT.
Using this contracted density matrix, the energy of the excited state
in the $\theta'$ representation for $H_{\rm{ref}}$ is given as
\begin{equation}
  E^{\rm{SC}}_{n_{\theta'}}
  = \frac{\left<\Xi_{n_{\theta'}} \right|
    H_{\rm{ref}} \left| \Xi_{n_{\theta'}} \right>}
  {\left. \left<\Xi_{n_{\theta'}} \right| \Xi_{n_{\theta'}} \right>}
  = E^{(0)}_{n_{\theta}} + \frac{\left<\Phi^{(0)}_{n_{\theta}} \right|
    V^{\dagger}_{\theta', \theta} \left[H_{\rm{ref}}, V_{\theta',
        \theta} \right] \left| \Phi^{(0)}_{n_{\theta}} \right>}
  {\left. \left<\Xi_{n_{\theta'}} \right| \Xi_{n_{\theta'}} \right>}
  \label{e_sc}
\end{equation}
where we have used Eqs.~\eqref{e0} and \eqref{xi} in the second
equality, and the strongly contracted approximation for the
correction,
$E^{(2):\rm{SC}}_{n_{\theta}} \simeq E^{(2)}_{n_{\theta}}$, becomes
\begin{equation}
  E^{(2):\rm{SC}}_{n_{\theta}} = \sum_{\theta' \neq \theta}
  \frac{\left. \left< \Xi_{n_{\theta'}} \right| \Xi_{n_{\theta'}}
    \right>}{E^{(0)}_{n_{\theta}} - E^{\rm{SC}}_{n_{\theta'}}}.
  \label{sc}
\end{equation}
This simplified approximation is possible because of Eq.~\eqref{xi},
which is a direct consequence of the optimal partitioning.  The
computational cost for the strongly contracted method is polynomial
and it does not require any diagonalization---we discuss more about
the scaling of the computational cost in
Sec.~\ref{sec:possible_issues}.

\subsection{$\mathbb{Z}_2$ symmetry-based PT and other multireference
  perturbation theories}
\label{sec:other_pt}

In the previous subsection, we have seen some similarities between
SBPT2 and NEVPT2.  We denote a second-order perturbation theory as
PT2.  In this section, we show that SBPT is in fact an extension of
other multireference perturbation theories (MRPTs) and argue why SBPT
could perform better in certain situations.

In the rest of this paper, we restrict ourselves to $\mathbb{Z}_2$
symmetries. We consider a set $L_a$ of spin orbitals where the number
of electrons in $L_a$ is either 0 or 1 modulo 2. The symmetry
operator, whether it comes from the $\mathbb{Z}_2$ subgroup of point
group symmetry or particle number symmetry, is then given by
\begin{equation}
  U_a = \prod_{p \in L_a} e^{i \pi \hat{N}_p} = \prod_{p \in L_a}
  \left(1 - 2 a^{\dagger}_p a_p \right)
  \label{z2_op}
\end{equation}
where $\hat{N}_p = a^{\dagger}_p a_p$ is the number operator for the
spin orbital $p$.  One could consider other abelian groups such as
$\mathbb{Z}_3$, but $\mathbb{Z}_2$ symmetry is the simplest and is
also of great interest in the context of quantum computing
applications as discussed in Sec.~\ref{sec:quantum_computing}.  For a
given Slater determinant, its character $e^{i\theta(a)}$ is $1$ if the
number of electrons in $L_a$ is even and $-1$ if it is odd. The
irreducible representation is uniquely determined by a set of
characters $\theta(a)$ on the generators $U_a$ of the group. Moreover,
the perturbation $V_{\theta', \theta}$ in Eq.~\eqref{v} can be easily
constructed for $\mathbb{Z}_2$: the operator $\mathcal{O}_{\rm{pert}}$
belongs to $V_{\theta', \theta}$ where $\theta'(a) = \theta(a)$ for
$[\mathcal{O}_{\rm{pert}}, U_a] = 0$ and
$e^{i\theta'(a)} = - e^{i\theta(a)}$ for
$[\mathcal{O}_{\rm{pert}}, U_a] \neq 0$.

\begin{figure}
  \centering
  \includegraphics[width=6cm]{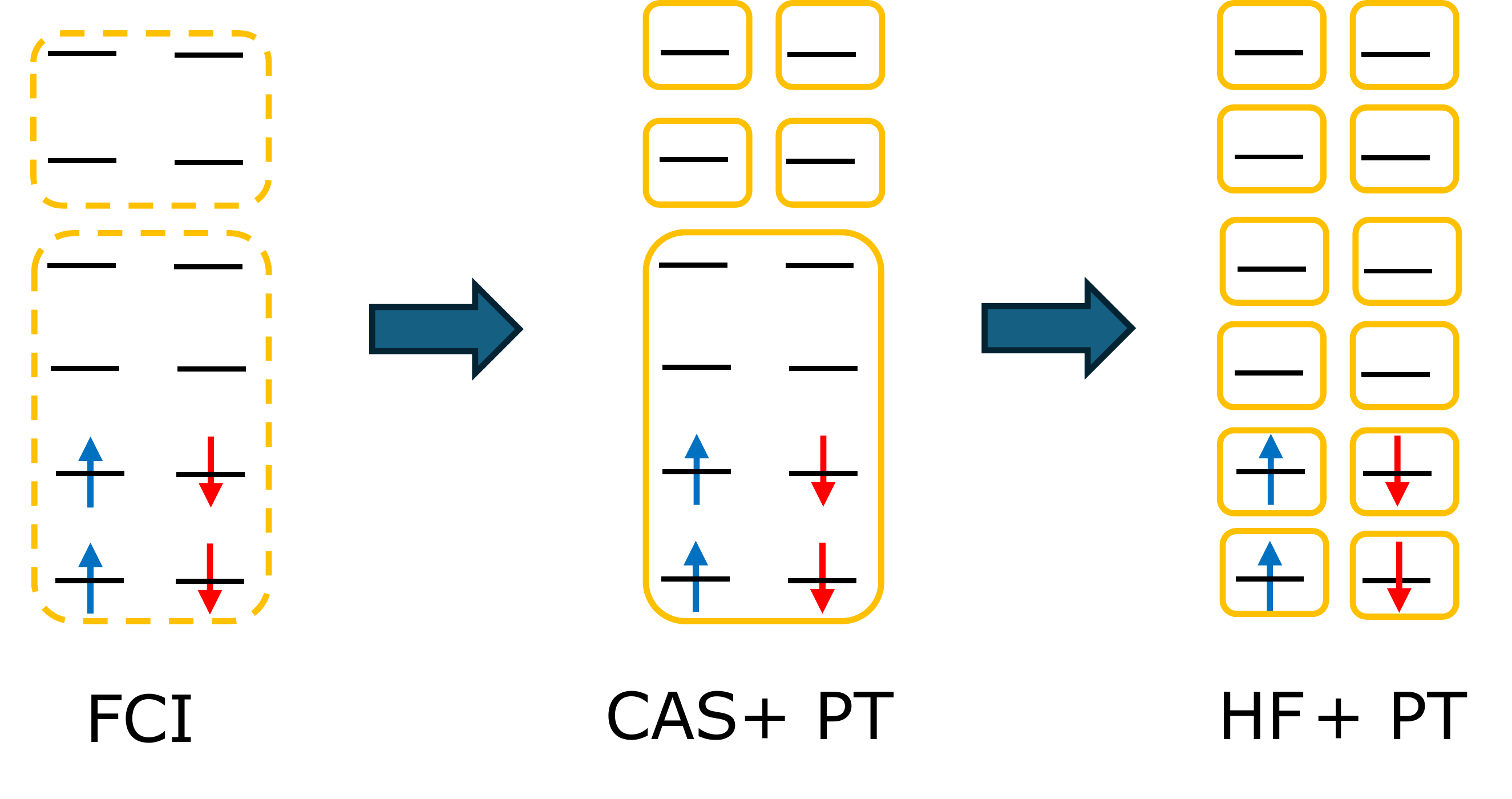}
  \caption{In this orbital energy diagram, the highest four spin
    orbitals couple weakly in FCI.  In the middle figure, the standard
    MRPTs treat them as external orbitals and form a CAS with the rest
    of the orbitals. In other words $H_{\rm{ref}}$ has $\mathbb{Z}_2$
    symmetries for each external spin orbital. The right figure
    corresponds to a situation where each spin orbital has
    $\mathbb{Z}_2$ symmetry, so that the HF state is the eigenstate of
    $H_{\rm{ref}}$. The computational cost decreases as we make more
    approximations.}
  \label{fig:cas_pt}
\end{figure}

We now argue that some existing MRPTs are SBPT with $\mathbb{Z}_2$
symmetry.  Let us assume that the four highest spin orbitals in a set
$L$ are weakly coupled, so that there is an approximate symmetry for
the number of electrons in $L$. We again use a dashed box to indicate
the approximate symmetry, as shown on the left in
Fig.~\ref{fig:cas_pt}.  Furthermore, we assume that each spin orbital
in $L$ does not couple much among themselves, and thus each has an
approximate $\mathbb{Z}_2$ symmetry. This corresponds to four sets
$L_a$ with $a = 1, 2, 3, 4$ as illustrated by the four small boxes in
the middle figure of Fig.~\ref{fig:cas_pt}. Performing SBPT for the
four $\mathbb{Z}_2$ symmetries, we have the Slater determinants in the
irreducible representation $\theta(a) = 0$ for all $a$ as unperturbed
wavefunctions, and the Slater determinants having a nonzero electron
number $\theta'(a) \neq 0$ for some $a$ are treated perturbatively.
This is equivalent to standard MRPTs, where we compute complete active
space (CAS) without the orbitals in $L_a$, i.e., the external
orbitals, and treat the external orbitals as perturbation. If the
Hartree-Fock state is the dominant contribution for the exact
solution, then we can perform a SBPT where each spin orbital has
$\mathbb{Z}_2$ symmetry, as illustrated on the right in
Fig.~\ref{fig:cas_pt}. This corresponds to a standard single-reference
perturbation theory (SRPT).

\begin{figure}
  \centering
  \includegraphics[width=10cm]{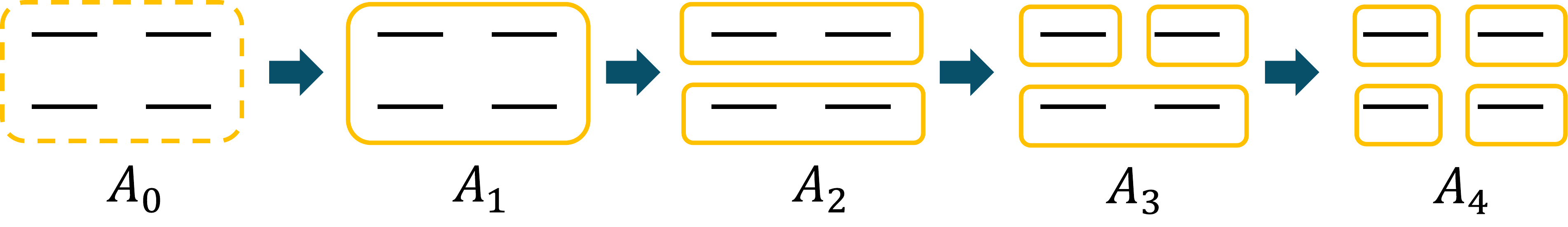}
  \caption{The four spin orbitals in $A_0$ couple
    nontrivially. Besides $A_4$ in standard MRPTs, SBPT can explore
    other groupings of $\mathbb{Z}_2$ symmetries, such as $A_1$,
    $A_2$, and $A_3$ shown here.}
  \label{fig:grouping}
\end{figure}

In fact, the SRPT based on the single Hartree-Fock (HF) state on the
right figure of Fig.~\ref{fig:cas_pt} with the optimal partitioning in
Eq.~\eqref{optimal_partitioning} corresponds to the Epstein-Nesbet
perturbation theory (ENPT).  Since each spin orbital has
$\mathbb{Z}_2$ symmetry, all the fermionic operators that destroy and
create a same spin orbital belong to the reference Hamiltonian. On the
other hand, the reference Hamiltonian of Moller-Plesset perturbation
theory (MPPT) only contains one-electron operators,
$H_{\rm{ref}} =\sum_{p} \epsilon_p a^{\dagger}_p a_p$, which respect
the same symmetry as ENPT, but this partitioning is not optimal due to
the lack of two-electron operators. However, MPPT is considered a more
reliable single-reference perturbation theory than ENPT, partly due to
its extensiveness and robust error cancellation \cite{Fink2016WhyDM}.

The MRPT based on CAS in the middle figure of Fig.~\ref{fig:cas_pt}
with the optimal partitioning in Eq.~\eqref{optimal_partitioning} is a
version of NEVPT discussed in
Ref.~\cite{Angeli2001IntroductionON}. Using their notation, the
reference Hamiltonian is given as
\begin{equation}
  H_{\rm{ref}} = \mathcal{P}_{\rm{CAS}} H
  \mathcal{P}_{\rm{CAS}} +  \sum_{l, k} \mathcal{P}_{S^{k}_{l}} H
  \mathcal{P}_{S^{k}_{l}}
  \label{nevpt_projection}
\end{equation}
where $\mathcal{P}_{S}$ is a projection operator onto the space $S$.
The index $k$ is the number of additional electrons in the active
space, and $l$ is the label for the electron configuration in the
inactive orbitals, i.e., the orbitals that are not in the active
space. Therefore their label $(k,l)$ has a one-to-one correspondence
with our label for the irrep $\theta'$, both of which specify electron
configurations in the inactive orbitals. If we employ the CAS and use
the Epstein-Nesbet approximation in Eq.~\eqref{en} for the
second-order correction, the reference Hamiltonian becomes
\begin{equation}
  H_{\rm{ref}} = \mathcal{P}_{\rm{CAS}} H
  \mathcal{P}_{\rm{CAS}} + \sum_{\theta'} \sum_{I_{\theta'}}
  E_{I_{\theta'}} \left| \Phi_{I_{\theta'}} \right> \left<
    \Phi_{I_{\theta'}} \right|.
\end{equation}
This reference Hamiltonian was mentioned in the original NEVPT paper
\cite{Angeli2001IntroductionON}, and it is sometimes called the
multireference Epstein-Nesbet perturbation theory in literature
\cite{Song_2020}.

The standard NEVPT uses a model called the Dyall Hamiltonian for
$H_{\rm{ref}}$, but it employs the same symmetry and the optimal
partitioning as the one in Eq.~\eqref{nevpt_projection}.  On the other
hand, CASPT uses the one-electron operator for $H_{\rm{ref}}$ similar
to MPPT. In this case, there is a well-known issue called the
intruder-state problem, where the energies of the perturber states are
nearly degenerate with the energy of the unperturbed state.  The
perturbative term then becomes very large or singular, leading to
divergent, oscillatory, or unstable PT
corrections~\cite{andersson1992second,roos1995multiconfigurational,
  forsberg1997multiconfiguration,battaglia2022regularized,
  sokolov2024multireference}.  This is partly because the two-electron
interaction is missing in the reference Hamiltonian. There are several
ways to mitigate the problem, such as using a level shift
\cite{roos1995multiconfigurational} or modifying the Hamiltonian
\cite{https://doi.org/10.1002/jcc.10098}. Our SBPT calculations are
based on the optimal partitioning, and therefore they include all one-
and two-electron interactions that respect the symmetry.  In the
examples we studied, we did not encounter any intruder-state problems.
However, both theoretical and numerical investigations into the
conditions for intruder-state problems in SBPT will be needed in
future studies. In short, we find that all SRPTs and MRPTs discussed
above can be seen as a version of $\mathbb{Z}_2$ SBPT with or without
the optimal partitioning.

Framing MRPT in the language of SBPT, we find that the method to
perform the standard MRPT may be restrictive, and there is a
straightforward way to extend it. Suppose again that four spin
orbitals are nontrivially coupled.  We denote this grouping as $A_0$
as shown in Fig.~\ref{fig:grouping}. In this example, existing MRPT
methods use the grouping $A_4$, where each spin orbital has a
$\mathbb{Z}_2$ symmetry as we discussed just above.  On the other
hand, there are many other ways to group the spin orbitals, depending
on how they are coupled among themselves. We illustrate some examples
of different grouping $A_n$ for $n=1,2,3$, where $n$ is the number of
$\mathbb{Z}_2$ symmetries. The larger the number $n$ of grouping, the
more symmetries we have to reduce the problem size. SBPT is therefore
a generalization of MRPT, and it extends and opens up this new window
of the grouping $A_n$ of $\mathbb{Z}_2$ symmetries. Even when the
standard MRPT using $A_4$ fails, the SBPT for different grouping $A_n$
with $n=1,2,3$ may give an accurate result. We demonstrate that this
indeed happens in Sec.~\ref{sec:examples}.

Lastly, we briefly mention another multireference perturbation theory
called symmetry-adapted perturbation theory (SAPT); see
Ref.~\cite{Jeziorski1994PerturbationTA} and references therein.  In
SAPT, one considers a weakly interacting dimer $AB$ and constructs
separate wave functions for monomers $A$ and $B$ at zeroth order. The
small intermolecular interaction is treated as a perturbative
correction, and the anti-symmetrization of the wavefunction is
explicitly enforced in the perturbative expansion.  Therefore, the
construction of the reference Hamiltonian in SAPT is not necessarily
guided by symmetry and is thus different from that in SBPT.

\subsection{Possible issues in SBPT2}
\label{sec:possible_issues}

In this section, we discuss some of the possible issues we may face in
SBPT2. First, as we increase the number of orbitals, there can be
factorially many ways to form a grouping $A_n$ for a given number of
symmetries.  Here we consider two methods to find a suitable grouping
that may give a good result. First, the chemistry can guide us to find
nontrivially coupled orbitals by using the symmetries and the
properties of orbitals. In general, finding a suitable grouping using
chemistry knowledge may not be straightforward, and this is a similar
situation as finding proper inactive orbitals in other standard MRPTs.
Second, we can choose a grouping that gives a smaller norm of the
perturbation Hamiltonian, $\lVert H_{\rm{pert}} \rVert$. This will
still give factorially many combinations, but it has an advantage of
automating the search without the knowledge of the molecule.  In
practice, we use two approaches together.  For example, if there are
several suitable groupings from chemistry reasoning, we can choose the
one that gives the smallest norm of $H_{\rm{pert}}$. We will study
this grouping for smaller molecular systems in Sec.~\ref{sec:examples}
as an initial investigation.

Another possible issue is the size consistency, which requires that
the energy of a dimer $AB$ equals the sum of the energies of the
monomers $A$ and $B$ in the dissociation limit. For example,
Moller-Plesset PT is known to be size consistent due to the linked
cluster theorem. We do not know whether NEVPT has a corresponding
linked cluster theorem for multireference formulations.  However, the
paper \cite{Angeli2001NelectronVS} explicitly shows that NEVPT
satisfies the size-consistency requirement for the Dyall
Hamiltonian. In this paper, we do not require the size consistency and
defer investigation of more general physical groupings that preserve
size consistency to future research.

Next, we provide a rough estimate of the computational cost for
SBPT2. We first choose a complete active space CAS($N$,$M$), where $N$
is the number of electrons and $M$ is the number of orbitals. We then
construct a reference Hamiltonian $H_{\rm{ref}}$ that possesses larger
$\mathbb{Z}_2$ symmetries with $K$ extra generators. The zeroth-order
eigenvalue calculation requires the diagonalization of the
$D$-dimensional $H_{\rm{ref}}$ matrix, where $D$ is the number of
determinants. Due to the augmented symmetry in $H_{\rm{ref}}$, the
number of determinants $\tilde{N}_{\rm{det}}$ in SBPT2 is smaller than
the number $N_{\rm{det}}$ in CAS($N$,$M$).  The actual number of
determinants depends on the symmetry in $H_{\rm{ref}}$, $M$, and $N$,
but we have $\tilde{N}_{\rm{det}} / N_{\rm{det}} = 2^{-K} $ at
best. The reason that it is typically less than $2^{-K}$ is because
the number $N_{\rm{det}}$ of determinants is already reduced by the
U(1) symmetries mentioned in
Sec.~\ref{sec:electronic_hamiltonian}. The computational cost of the
diagonalization scales as $\mathcal{O}(M^4 N_{\rm{det}})$ per
iteration in Davidson's method, where $M^4$ comes from the number of
fermionic operators in the reference Hamiltonian. In SBPT, both the
number of fermionic operators and the number of determinants are
smaller due to additional symmetries. Therefore the leading-order
calculation always favors SBPT2 over other MRPTs based on CAS. This
advantage manifests as a reduction in qubit counts in the quantum
computing application as explained in
Sec.~\ref{sec:quantum_computing}.

At second order, the majority of the computational cost in the
strongly contracted method comes from the computation of
Eq.~\eqref{e_sc}. In NEVPT2, it requires the computational cost of the
4-RDM, which scales as $\mathcal{O} (M^8 N_{\rm{det}})$, rather than
the 6-RDM due to the commutation in Eq.~\eqref{e_sc}
\cite{Angeli2001NelectronVS,
  Angeli2002nelectronVS,sokolov2024multireference}. For the same
fermionic operators in $H_{\rm{pert}}$, we expect the same scaling but
with a lower cost because the number of determinants is smaller, and
the eight indices in the 4-RDM are constrained by the symmetries.

There are, however, additional terms in $H_{\rm{pert}}$ due to the
augmented symmetry. Consider a set $L_a$ of spin orbitals that
transform nontrivially under one extra generator.  The number of extra
terms in $H_{\rm{pert}}$ due to this extra generator scales as
$M_a M^3$ where $M_a$ is the number of orbitals in $L_a$.  There are
then $\sim M^2_a M^8$ operators to compute in Eq.~\eqref{e_sc} after
using the commutation. Again the number should be much smaller because
the indices are constrained by the symmetries. For $K$ extra
generators, the total extra cost scales as
$\mathcal{O}(\max(M_a)^2 K M^8 \tilde{N}_{\rm{det}})$.  This
additional cost does not become a bottleneck as long as we have
$\max(M_a)^2 K \tilde{N}_{\rm{det}} / N_{\rm{det}} \lesssim 1$. In
general, we assume that $M_a \ll M$, or else the number $K$ of extra
symmetries becomes negligible compared to $M$. If $\max(M_a) \sim 1$,
as we have in Sec.~\ref{sec:examples}, then the condition
$K \tilde{N}_{\rm{det}} / N_{\rm{det}} \lesssim 1$ should hold for any
$K$.

Lastly we mention that the effectiveness of SBPT diminishes as the
active space size $M$ increases for a fixed number of symmetries. We
will study this issue for the case of $\mathrm{N_2}$ using two
different basis set sizes in Sec.~\ref{sec:n2} and show that the
number of symmetries can increase as we increase the basis size. We
defer a more thorough investigation of all the issues above to future
research.

\section{Applications of SBPT}
\label{sec:applications}

In this section, we consider two applications of SBPT. First we
implement selected configuration interaction (SCI) in SBPT to overcome
the shortcomings of perturbation theory. In the following section, we
show that our method can be incorporated naturally into quantum
computing applications to reduce the number of qubits using the
so-called qubit tapering technique.

\subsection{SCI-SBPT}
\label{sec:sci_sbpt}

The perturbative expansion for electronic structure calculations is an
approximate method, which is typically neither convergent nor
variational.  A higher-order correction may result in a large error
and/or a lower value than the true ground-state energy. To circumvent
these problems, we can employ selected configuration interaction (SCI)
~\cite{tubman2016deterministic,holmes2016heat,
  10.1063/1.4948308,zhang2025random} wherein the ground state is
approximated by a sparse linear combination of electronic
configurations~\cite{bender1969studies,huron1973iterative,
  buenker1974individualized,buenker1978applicability,
  evangelisti1983convergence,illas1991selected,harrison1991approximating,
  ivanic2001identification,stampfuss2005improved,bytautas2009priori,
  roth2009importance}.

The SCI selects important configurations and diagonalizes the matrix
exactly within the selected subspace. A crucial aspect of SCI is to
identify which configurations are most important.  One of the common
approaches is to use the second-order correction of Moller-Plesset
perturbation theory as a selection criterion. This method can be
naturally extended to SBPT.

We consider the strongly contracted approximation of the second-order
correction given in Eq.~\eqref{sc}:
\begin{equation}
  E^{(2):SC}_{n_{\theta}} = \sum_{\theta' \neq \theta}
  \frac{\left. \left< \Xi_{n_{\theta'}} \right| \Xi_{n_{\theta'}}
    \right>}{E^{(0)}_{n_{\theta}} - E^{SC}_{n_{\theta'}}}
  = \sum_{\theta' \neq \theta} E_{\theta'}.
\end{equation}
There are many possible ways to select important configurations using
this second-order correction, but a straightforward approach is to
select irreducible representations $\theta'$ with
\begin{equation}
  \left| \frac{E_{\theta'}}{E^{(0)}_{n_{\theta}}} \right| > \epsilon_1,
  \label{epsilon_1}
\end{equation}
where $\epsilon_1$ is a small cutoff to regulate the number of
irreps. We denote the number of irreps in SCI as $N_{\theta}$, which
is determined by $\epsilon_1$.  For the state in the selected irreps,
\begin{equation}
  \left| \Xi_{n_{\theta'}} \right> = \sum_{I_{\theta'}}
  c_{I_{\theta'}} \left| \Phi_{I_{\theta'}} \right>,
\end{equation}
we select the configurations with
$\left| c_{I_{\theta'}} \right| > 0$.

We can further select the configurations by omitting those that
contribute little to the state $\left| \Xi_{n_{\theta'}}
\right>$. That is, for the state in the selected irrep,
$\left| \Xi_{n_{\theta'}} \right>$ that satisfies
Eq.~\eqref{epsilon_1}, we choose the configurations with a coefficient
larger than some cutoff $\epsilon_2$:
\begin{equation}
  \left| c_{I_{\theta'}} \right| > \epsilon_2 .
  \label{epsilon_2}
\end{equation}
This condition is familiar from other calculations, including exact
diagonalization using Davidson's method.

We can tweak these two parameters to select the number of
configurations to fit within available computational resources. While
one could vary $\epsilon_2$ depending on the value of $\epsilon_1$, we
consider $\epsilon_1$ and $\epsilon_2$ as fixed parameters. We make
the selection criterion as simple as possible in this paper.

\subsection{Quantum computing applications}
\label{sec:quantum_computing}

Symmetry-based perturbation theory can be straightforwardly
implemented in quantum computing applications. Once we obtain the
second quantized Hamiltonian Eq.~\eqref{2nd_q}, we can transform it in
terms of Pauli strings,
$P_{\alpha} \in \left\{I, X, Y, Z \right\}^{\otimes N_Q}$:
\begin{equation}
  H = \sum^{N_{P}}_{\alpha = 1} c_{\alpha} P_{\alpha}.
\end{equation}
The qubit Hamiltonian can then be simulated on quantum computers.  The
mapping to the qubit Hamiltonian is not unique and there are several
methods. We consider the standard Jordan-Wigner transformation here,
where the number of qubits $N_Q$ equals the number of spin orbitals
$2M$, and there is one-to-one correspondence between the occupation
number states of a spin orbital and the states of a qubit.

The $\mathbb{Z}_2$ symmetry operator in Eq.~\eqref{z2_op} can be
written in the Jordan-Wigner transformation as
\begin{equation}
  U_a = \prod_{p \in L_a} Z_p.
    \label{generator}
\end{equation}
$Z_p$ is a Pauli string where it is $Z$ for the $p$-th qubit and $I$
for the others. We consider $K$ generators of the abelian group, $U_a$
with $a = 1, \dots, K$. Since $[P_\alpha, U_a] = 0$ for all $\alpha$
and any $U_a$ from Eq.~\eqref{fop_commutation}, we can construct a
Clifford operator $C$ using the $K$ generators and perform the
Clifford transformation on the Hamiltonian
\cite{Bravyi2017TaperingOQ}:
\begin{equation}
  H \rightarrow CHC^{\dagger} = \sum^{N_{P}}_{\alpha = 1} c_{\alpha}
  Q_{\alpha}
\end{equation}
where $Q_{\alpha} = C P_{\alpha} C^{\dagger}$ is a Pauli string with
an overall $+$ or $-$ sign.  Since the generator is in the $Z$
operator basis as given in Eq.~\eqref{generator}, we can show that
there are $K$ different $X_{q(i)}$ operators with $i = 1, 2, \dots, K$
for qubit index $q(i)$ such that
$\left[Q_{\alpha}, X_{q(i)} \right] = 0$ for all $\alpha$. This means
that the eigenstate in Eq.~\eqref{eigenstate} in the new basis,
$C\left| \Psi_{n_{\theta}} \right>$, is also an eigenstate of
$X_{q(i)}$ for $i = 1, 2, \dots, K$, which can be uniquely determined
by the irreducible representation $\theta$ of the state. Thus we can
effectively reduce the number of qubits from $N_Q$ to $N_Q - K$. This
is called qubit tapering \cite{Bravyi2017TaperingOQ}. The relation of
qubit tapering with the point group symmetry is further discussed in
Ref.~\cite{doi:10.1021/acs.jctc.0c00113} and its application to
chemistry problems is discussed in
Refs.~\cite{doi:10.1021/acs.jctc.3c00228, Picozzi_2023}.

If other transformations to the qubit Hamiltonian are related to the
Jordan-Wigner transformation by a Clifford transformation, such as the
Bravi-Kitaev transformation, then the number of qubit tapering remains
the same.  For $\mathbb{Z}_N$ symmetries with $N>2$, the symmetry
operator in the Jordan-Wigner transformation is in general written as
a linear combination of multiple Pauli strings, and it cannot be used
for qubit tapering in a way described above.

There are at least two $\mathbb{Z}_2$ symmetries associated with the
spin and up to three additional $\mathbb{Z}_2$ symmetries associated
with the exact point-group symmetry. The qubit tapering with the two
$\mathbb{Z}_2$ symmetries of spin is associated with the parity
mapping, although it gives rise to two different Hamiltonians. The
$\mathbb{Z}_2$ subgroups of the point group symmetry are the
inversions and $180$-degree rotations.  Since there are three
coordinates, $\mathbf{x} = (x,y,z)$, we can have at most three
independent $\mathbb{Z}_2$ transformations, such as
$\mathbf{x}_i \rightarrow - \mathbf{x}_i$ for each $i$.  Therefore
$K=5$ is the maximum number possible in the electronic Hamiltonian.

Using SBPT, we can increase the number $K$ of $\mathbb{Z}_2$
symmetries to any number up to $N_Q$.  We consider additional
generator $U_{b}$ with $[H_{\rm{ref}}, U_b] = 0$ but
$[H, U_b] \neq 0$, where $L_b$ is the set of spin orbitals that are
nontrivially coupled as discussed in
Sec.~\ref{sec:approximate_symmetries}. Assuming that we have $K'$ such
additional generators, $U_b$ with $b = 1, \dots, K'$, we can show that
the optimal partitioning Eq.~\eqref{optimal_partitioning} can be
translated into the Pauli string representation as
\begin{equation}
\left[ P_{\alpha}, U_b \right] \neq 0
\end{equation}
for some $U_b$ and for all $P_{\alpha}$ in $H_{\rm{pert}}$. Again this
optimal partitioning uniquely determines the reference Hamiltonian, as
well as $V_{\theta',\theta}$ in Eq.~\eqref{v}.  The reference
Hamiltonian has $K'$ additional generators, so we can reduce
additional $K'$ qubits.

SCI-SBPT2 in the previous section is also relevant in quantum
computing applications.  The use of selected CI in quantum computing
applications was first proposed in Ref.~\cite{dmn4-snfx} and
was called quantum selected configuration interaction (QSCI). The SQD
paper \cite{robledo2024chemistry} extended QSCI and employed a
recovery method, increasing the size of quantum circuits significantly
on noisy quantum hardware. While SCI-SBPT2 can be implemented
straightforwardly in QSCI, the recovery method in SQD needs
modification since the number of particles is not well defined. The
same issue arises for the parity mapping. However, SCI-SBPT2 without
qubit tapering can be used in both QSCI and SQD, and it can reduce the
number of configurations in $H_{\rm{ref}}$ to select from samples
obtained from the quantum hardware run.

\section{Results}
\label{sec:examples}

In this section, we apply second-order symmetry-based perturbation
theory (SBPT2) to the ground-state potential energy curves of two
molecules, $\mathrm{H_2O}$ and $\mathrm{N_2}$, in the STO-3G
basis. Through the examples, we show how SBPT2 works for systems with
different point-group symmetries and demonstrate that SBPT2 can give
accurate results with a lower computational cost than second-order
$n$-electron valence state perturbation theory (NEVPT2). We measure
the computational cost by the number of configurations and the number
of qubits required to solve for $H_{\rm{ref}}$. We also discuss
$\mathrm{N_2}$ in the 6-31G basis at the end.

We obtain the restricted Hartree-Fock orbitals with point-group
symmetries using the free open-source chemistry software PySCF
\cite{https://doi.org/10.1002/wcms.1340}. We have used PySCF to
compute the exact energy dissociation curves, second-order
Moller-Plesset perturbation theory (MP2), as well as the standard
NEVPT2 calculations based on the Dyall Hamiltonian with the strongly
contracted method. We denote NEVPT2 with CAS($N$,$M$) as
NEVPT2($N$,$M$), where $N$ is the number of electrons and $M$ the
number of orbitals.

For the other calculations, we have used the Jordan-Wigner
transformation of the second-quantized Hamiltonian in terms of Pauli
strings as in Sec.~\ref{sec:quantum_computing}. As discussed above, we
can use other transformations or the second quantized Hamiltonian
directly, and the results will be independent of the basis. For the
SBPT calculations, including a version of NEVPT2 based on the
projection Hamiltonian, we have used the Epstein-Nesbet (EN) and
strongly contracted (SC) approximations given in Eqs.~\eqref{en} and
\eqref{sc}, as well as the full uncontracted (UC) second-order
correction \eqref{2nd_order}. The UC correction is not scalable as
discussed above, but the purpose of showing it for the small systems
is to validate the SC approximation.  The SCI-SBPT2 calculation is
based on the SC approximation of the second-order correction.

\subsection{$\mathrm{H_2O}$}
\label{sec:h2o}

\begin{figure}
  \centering
  \begin{subfigure}{0.48\textwidth}
    \centering
    \includegraphics[width=\textwidth]{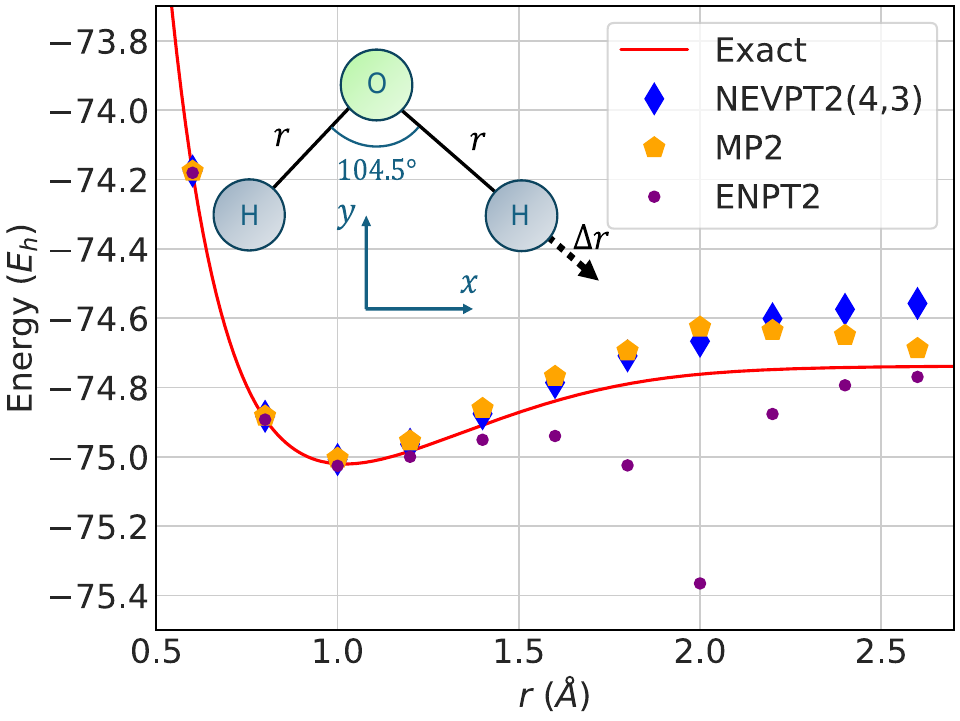}
    \caption{Dissociation curve}
    \label{fig:h2o_dissociation}
  \end{subfigure}
  \hfill
  \centering
  \begin{subfigure}{0.48\textwidth}
    \centering
    \includegraphics[width=\textwidth]{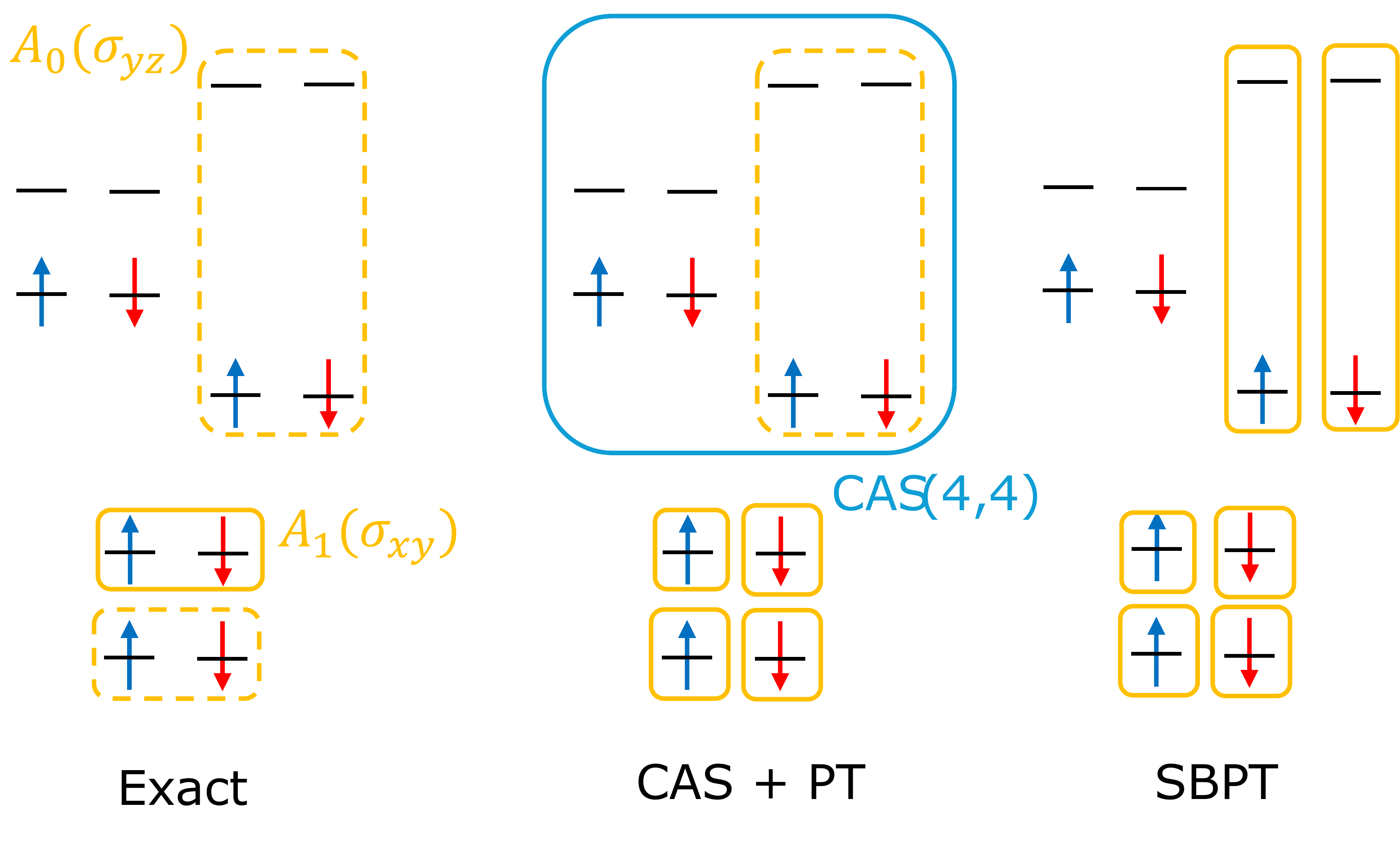}
    \caption{Orbital energy diagrams in MRPTs}
    \label{fig:h2o}
  \end{subfigure}
  \vfill
  \centering
  \begin{subfigure}{0.48\textwidth}
    \centering
    \includegraphics[width=\textwidth]{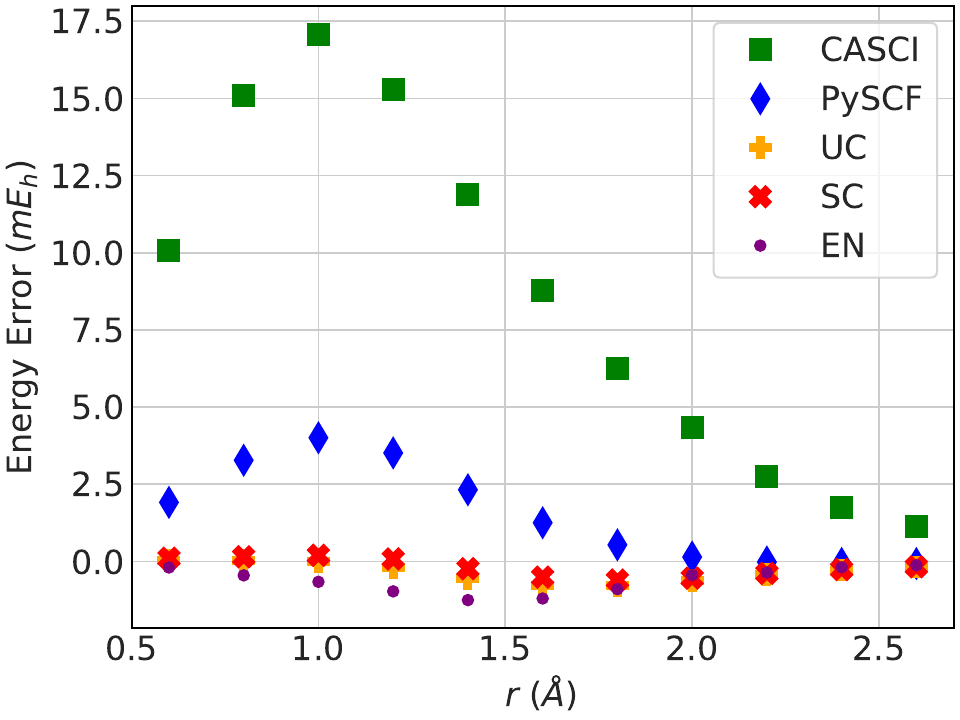}
    \caption{NEVPT2(4,4)}
    \label{fig:h2o_nevpt}
  \end{subfigure}
  \hfill
  \centering
  \begin{subfigure}{0.48\textwidth}
    \centering
    \includegraphics[width=\textwidth]{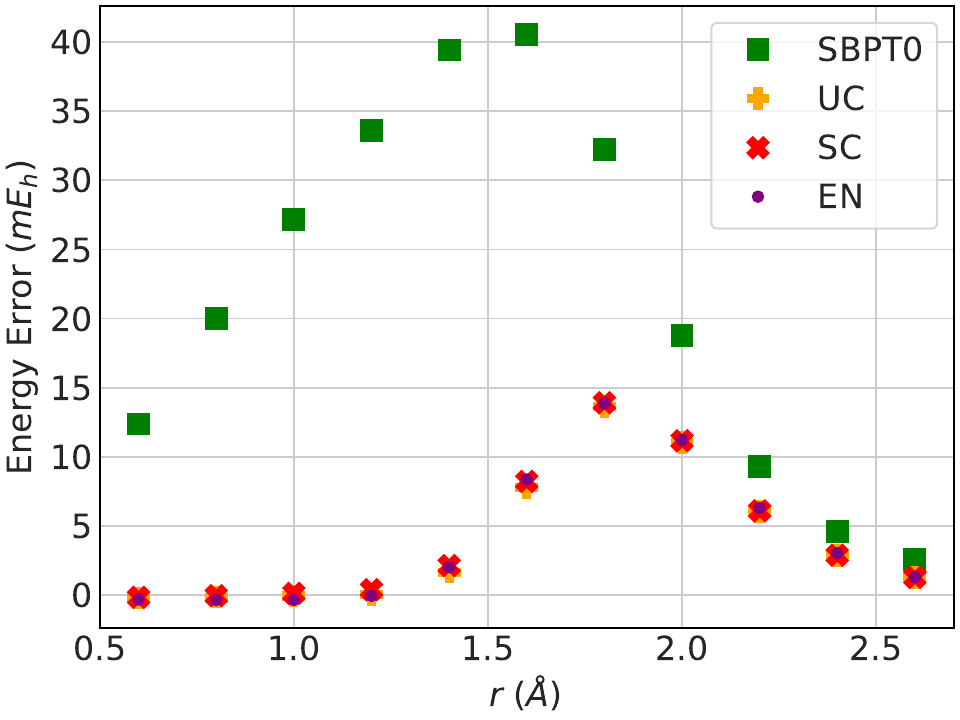}
    \caption{SBPT2}
    \label{fig:h2o_sbpt}
  \end{subfigure}
  \vfill
  \centering
  \begin{subfigure}[b]{0.48\textwidth}
    \centering
    \includegraphics[width=\textwidth]{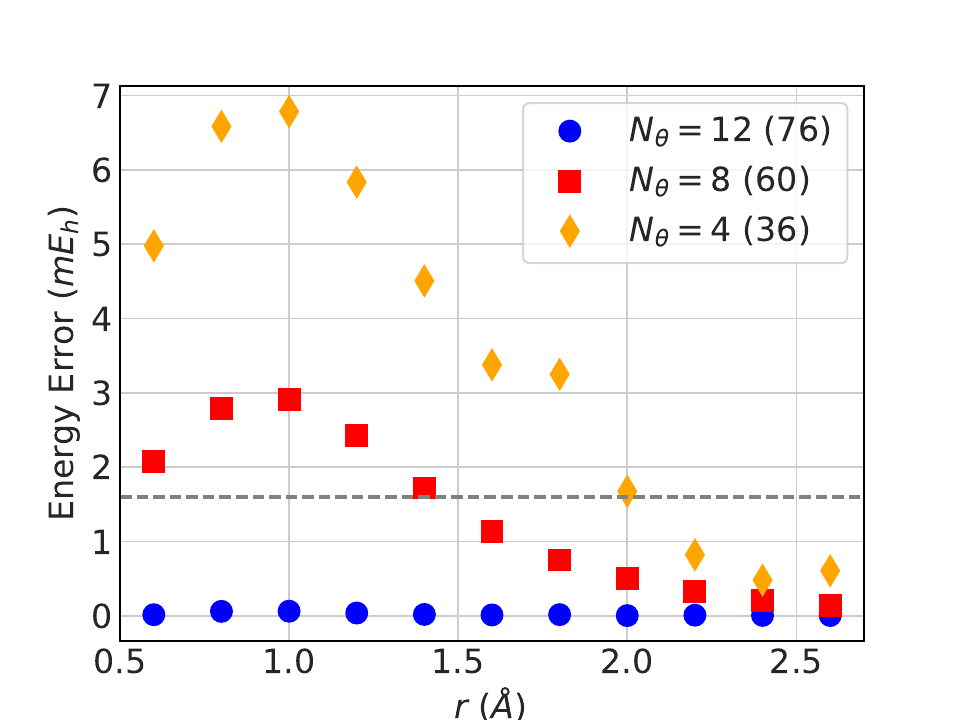}
    \caption{SCI-SBPT2: $\epsilon_2 = 0$}
    \label{fig:h2o_epsilon1}
  \end{subfigure}
  \hfill
  \centering
  \begin{subfigure}[b]{0.48\textwidth}
    \centering
    \includegraphics[width=\textwidth]{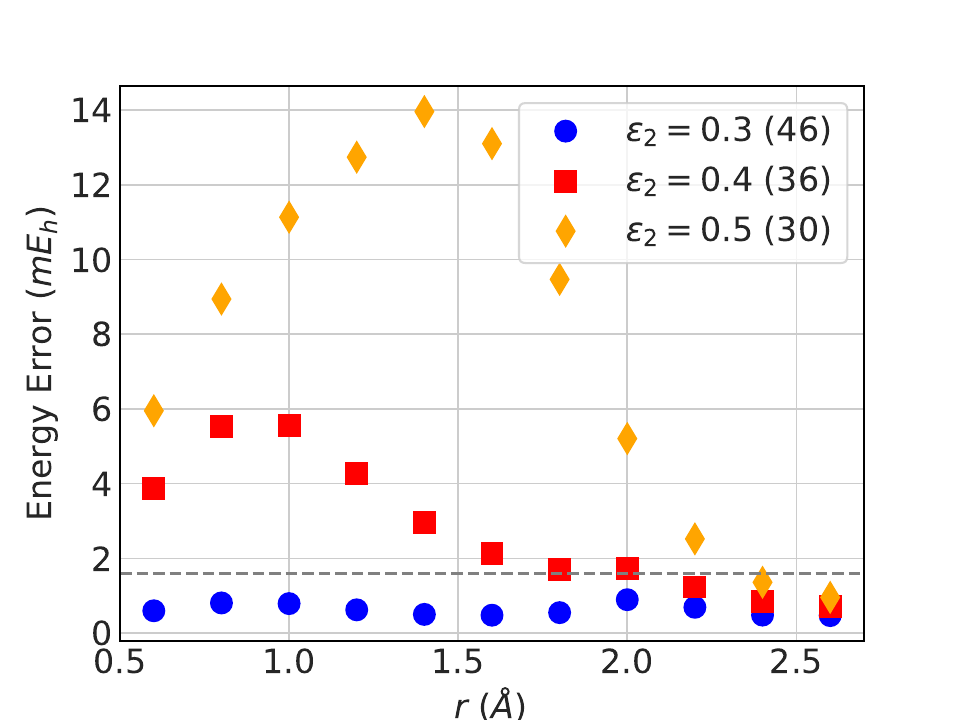}
    \caption{SCI-SBPT2: $N_{\theta} = 12$}
    \label{fig:h2o_epsilon2}
  \end{subfigure}
  \caption{Ground-state potential energy curve of $\mathrm{H_2O}$ in
    the STO-3G basis. (a) The dissociation curve for the deformed
    geometry, $\Delta r = 0.01$. (b) Schematic orbital energy diagrams
    for $r = 1.8$ with the groupings of the exact (solid) and
    approximate (dashed) $\mathbb{Z}_2$ symmetries in MRPTs. (c, d)
    NEVPT2(4,4) and SBPT2 results of the leading-order contributions
    and the second order corrections: the uncontracted (UC), strongly
    contracted (SC), and Epstein-Nesbet (EN) approximations are given
    in Eqs.~\eqref{2nd_order}, \eqref{sc}, and \eqref{en},
    respectively. PySCF uses the Dyall Hamiltonian with the SC
    approximation. (e, f) The SCI-SBPT2 results using the cutoffs
    given in Eqs.~\eqref{epsilon_1} and \eqref{epsilon_2}: varying
    $\epsilon_1$ with fixed $\epsilon_2=0$ and varying $\epsilon_2$
    with fixed $\epsilon_1$, which gives the number of irreps,
    $N_{\theta} = 12$. The number in parentheses is the number of
    selected configurations.}
  \label{fig:h2o_energies}
\end{figure}

We start with the ground state energy dissociation curve for the water
molecule in the STO-3G basis. We keep the angle between the two
hydrogen atoms at $104.5^{\circ}$ and stretch the two $\mathrm{O-H}$
bonds.  This ensures that the ground state has multireference
character at large bond distances. To demonstrate the use of
approximate point-group symmetries discussed in
Sec.~\ref{sec:approximate_symmetries}, we stretch one of the
$\mathrm{O-H}$ bonds slightly more than the other: one bond length is
set to be $r$ while the other is set to be $r + \Delta r$ with a small
deformation $\Delta r = 0.01$. We use the angstrom as the unit of
length in this paper.

In the STO-3G basis with one frozen core, there are six orbitals and
eight electrons. The frozen-core approximation for the energy
dissociation curve, which we treat as the exact solution, is given in
Fig.~\ref{fig:h2o_dissociation}. We compute various perturbative
approximations of the energy at 11 values of $r$ equally spaced from
$r = 0.6$ to $2.6$. Figure \ref{fig:h2o_dissociation} shows that the
NEVPT2(4,3), MP2, and second-order Epstein-Nesbet perturbation theory
(ENPT2) results have good agreement near the equilibrium geometry, but
they deviate significantly from the exact solution both at large bond
lengths as the system exhibits the multireference nature and at the
dissociation limit. This plot shows that single-reference perturbation
theory (SRPT) or multireference perturbation theory (MRPT) with small
active space cannot correctly compute the dissociation curve.

The point group symmetry of the $\mathrm{H_2O}$ molecule without the
deformation, $\Delta r = 0 $, is $C_{2v}$. There are two
$\mathbb{Z}_2$ subgroups associated with the reflections $\sigma_{xy}$
and $\sigma_{yz}$.  One of them, $\sigma_{yz}$, is explicitly broken
in the presence of the deformation $\Delta r \neq 0$. It is, however,
an approximate symmetry when $\Delta r$ is sufficiently small.  We
group the orbitals that transform nontrivially under the exact
symmetry $\sigma_{xy}$ in the solid box, and the approximate symmetry
$\sigma_{yz}$, as well as the approximate $\mathbb{Z}_2$ symmetry for
the lowest orbital that is weakly coupled, in the dotted boxes as
shown in Fig.~\ref{fig:h2o}.  There are multiple energy level
crossings of the orbitals, and the orbital energy diagram is for
$r = 1.8$.  In the exact case, the number of configurations that
respect the $\mathbb{Z}_2$ reflection symmetry associated with
$\sigma_{xy}$ and the two spin symmetries is 125. Since there are six
orbitals, the number of qubits in the Jordan-Wigner transformation
gives rise to 12 qubits.  Using the three $\mathbb{Z}_2$ symmetries,
we can reduce the number of qubits to 9. The summary of the resource
requirements is given in Table \ref{table:h2o}.

\begin{table}[h!]
  \centering
  \begin{tabular}{|c || c c c c|}
    \hline
    & Exact & CAS(4,4) & SBPT & CAS(4,3)\\ [0.5ex]
    \hline\hline
    Number of orbitals ($M$) & 6 & 4 & 4 & 3 \\
    \hline
    Number of qubits ($N_Q$) & 9 & 6 & 4 & 4 \\
    \hline
    Number of configurations ($N_{\rm{det}}$) & 125 & 36 & 16 & 9 \\ [1ex]
    \hline
  \end{tabular}
  \caption{The resource requirements---the number of qubits and
    configurations---for $\mathrm{H_2O}$ with the deformation in the
    STO-3G basis for the Hamiltonians defined in
    Fig.~\ref{fig:h2o}. The exact case is the frozen core
    approximation, i.e., CAS(8,6). CAS(4,3) is for reference
    purposes. The table is ordered by increasing number of
    symmetries.}
  \label{table:h2o}
\end{table}

We now look for a good perturbative approximation of the dissociation
curve with as few resources as possible. In other words, we aim to
find a good reference Hamiltonian with the largest number of
$\mathbb{Z}_2$ symmetries.

We can start with the conventional valence space to form the complete
active space, CAS(4,4), and treat the two lowest orbitals as external
orbitals as they are weakly coupled.  This is the standard MRPT
approach based on CASCI, and its orbital energy diagram with the
$\mathbb{Z}_2$ symmetries is shown in the middle figure of
Fig.~\ref{fig:h2o}. Figure \ref{fig:h2o_nevpt} shows that the
NEVPT2(4,4) results with the different approximations for the
second-order correction give good agreement with the exact solution.
We have explicitly checked that CAS(4,4) is the smallest active space
to obtain reasonable agreement with the exact solution. This is
expected, as we need to include all the $2p$ orbitals to keep the
rotational symmetry in the dissociation limit.  In CAS(4,4), the
number of configurations that respect the $\mathbb{Z}_2$ symmetries is
36, and the number of qubits after the qubit tapering is 6.

As just seen above, the standard MRPT approaches miss the opportunity
to exploit the approximate symmetry of $\sigma_{yz}$, while SBPT can
readily incorporate it in the reference Hamiltonian as the following.
The two orbitals $\phi_p$ in the dashed box in CAS(4,4) transform
nontrivially under the reflection in the $y-z$ plane,
$\phi_p \rightarrow -\phi_p + \Delta \phi_p$ where
$|\Delta \phi_p | \ll \left| \phi_p \right|$.  The deviation
$\Delta \phi_p$ goes to zero as we take $\Delta r$ to be zero. We
consider the reference Hamiltonian that gives this approximate
symmetry exact. This corresponds to the replacement of the dashed box
with a solid box, resulting in the reduction of one extra qubit.
Since the two orbitals in $A_0 (\sigma_{yz})$ are weakly coupled to
the rest of the orbitals, we can further decompose the exact symmetry
as shown in the right figure of Fig.~\ref{fig:h2o}. This corresponds
to the reduction of two extra qubits in total, which is the same
amount of reduction as CAS(4,3). Alternatively, we could divide $A_0$
as $A_2$ in Fig.~\ref{fig:grouping}, but this would give bad results
at large bond lengths for two reasons.  First the norm of the
perturbation in the alternative grouping is larger, and second the
grouping does not allow the spin singlet configurations, which are
important in the dissociation limit.  Therefore both the norm and
chemistry arguments can correctly guide us to choose a better
reference Hamiltonian in this example.

We find from Fig.~\ref{fig:h2o_sbpt} that the SBPT2 approximations
agree well with the exact solution at all bond lengths.  We also
observe that the SC and EN approximations give the same results as the
UC correction. This suggests that while the unperturbed state requires
the reference Hamiltonian that incorporates multireference nature,
each excited state in the reference Hamiltonian can be approximated in
the mean-field approach with a single Slater determinant. In the
intermediate range of the bond lengths where the correlation becomes
strong, we see larger energy errors, but they are comparable to
NEVPT2(4,4).  In summary, SBPT2 and NEVPT2(4,4) both give good
approximations, but SBPT2 can perform with a smaller computational
cost than NEVPT2(4,4) by exploiting the approximate reflection
symmetry.

As discussed in Sec.~\ref{sec:sci_sbpt}, all perturbative results are
neither variational, as shown in Fig.~\ref{fig:h2o_energies}, nor
convergent, even if we perform higher-order corrections.  To achieve
more accurate and systematic results, we could turn to SCI-SBPT2,
where the results are shown in Figs.~\ref{fig:h2o_epsilon1} and
\ref{fig:h2o_epsilon2} with the dotted line indicating the kcal/mol
accuracy, $1.6 \, mE_h$, as a reference point. We select
configurations at $r=1.8$ and use the same configurations for all the
other values of $r$ for simplicity.  There are 125 configurations that
respect the exact $\mathbb{Z}_2$ symmetries.  The exact
diagonalization of the ground state energy shows that we only need at
most 30 out of 125 configurations to achieve the accuracy.  With the
augmented $\mathbb{Z}_2$ symmetry in SBPT, these 125 configurations
belong to 25 different irreducible representations.  With zero cutoffs
$\epsilon_1 = \epsilon_2 = 0$, SCI-SBPT2 selects 116 configurations in
20 irreps. Figure \ref{fig:h2o_epsilon1} shows that at most 12 out of
20 irreps are relevant for the dissociation curve calculation.
Moreover, we see from Fig.~\ref{fig:h2o_epsilon2} that only 36
configurations in the 12 irreps can achieve the accuracy at $r = 1.8$,
which is comparable to the required number of configurations from the
exact diagonalization.

\subsection{$\mathrm{N_2}$}
\label{sec:n2}

\begin{figure}
  \centering
  \begin{subfigure}{0.48\linewidth}
    \centering
    \includegraphics[width=\linewidth]{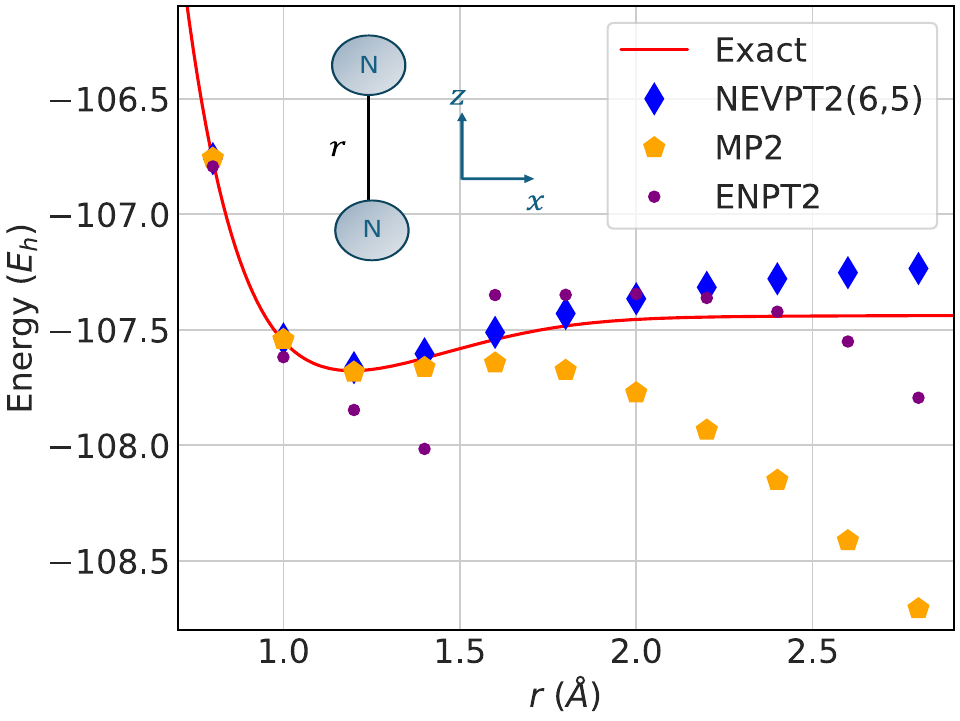}
    \caption{Dissociation curve}
    \label{fig:n2_dissociation}
  \end{subfigure}
  \hfill
  \centering
  \begin{subfigure}{0.48\linewidth}
    \centering
    \includegraphics[width=\linewidth]{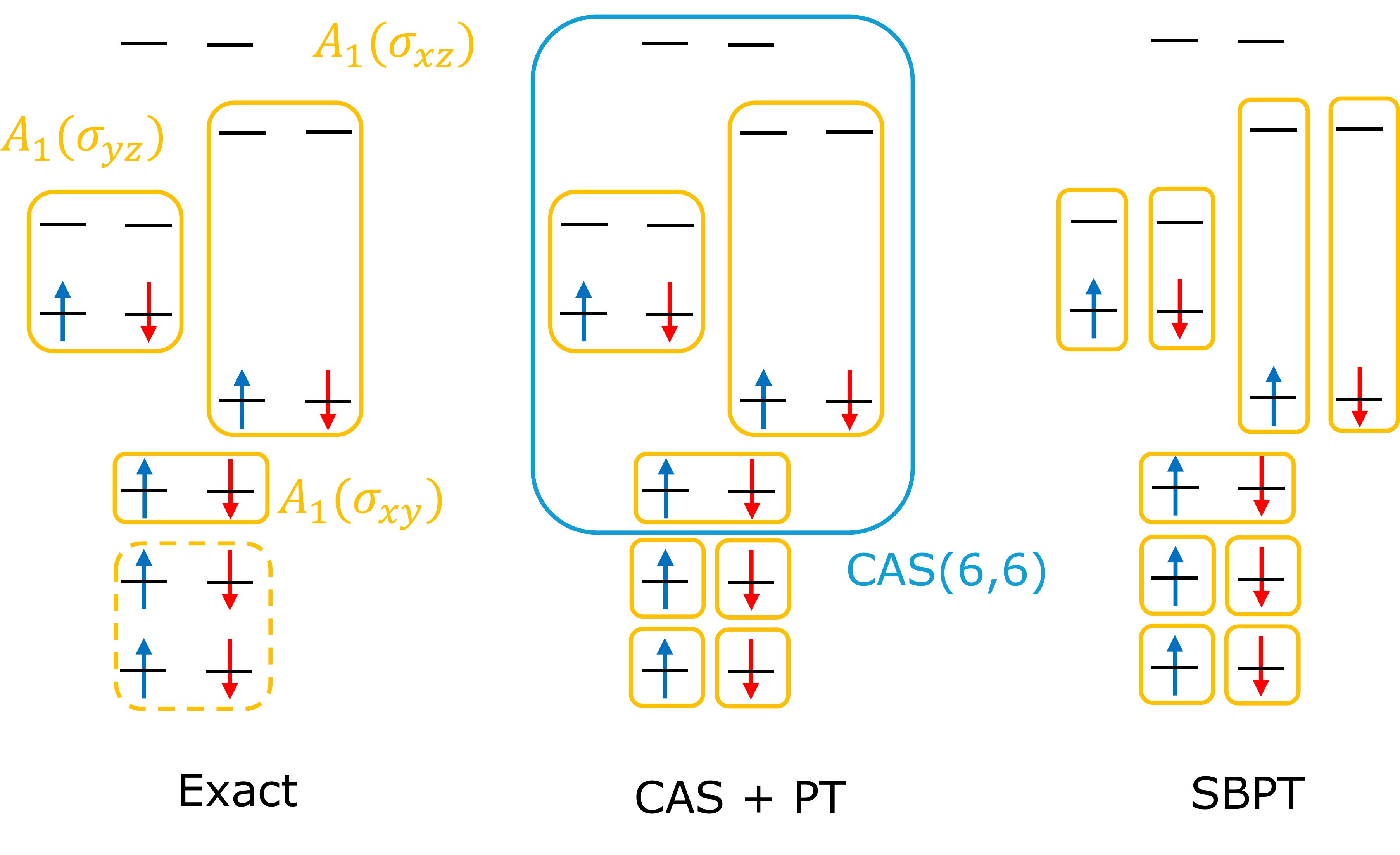}
    \caption{Orbital energy diagrams in MRPTs}
    \label{fig:n2}
  \end{subfigure}
  \vfill
  \centering
  \begin{subfigure}{0.48\linewidth}
    \centering
    \includegraphics[width=\linewidth]{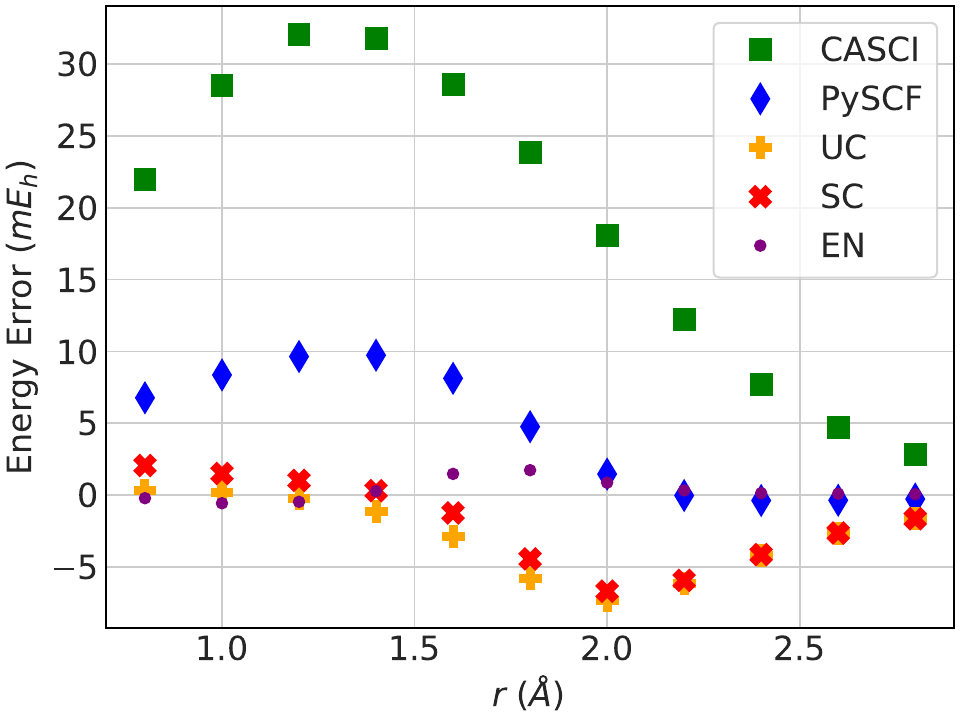}
    \caption{NEVPT2(6,6)}
    \label{fig:n2_nevpt}
  \end{subfigure}
  \hfill
  \centering
  \begin{subfigure}{0.48\linewidth}
    \centering
    \includegraphics[width=\linewidth]{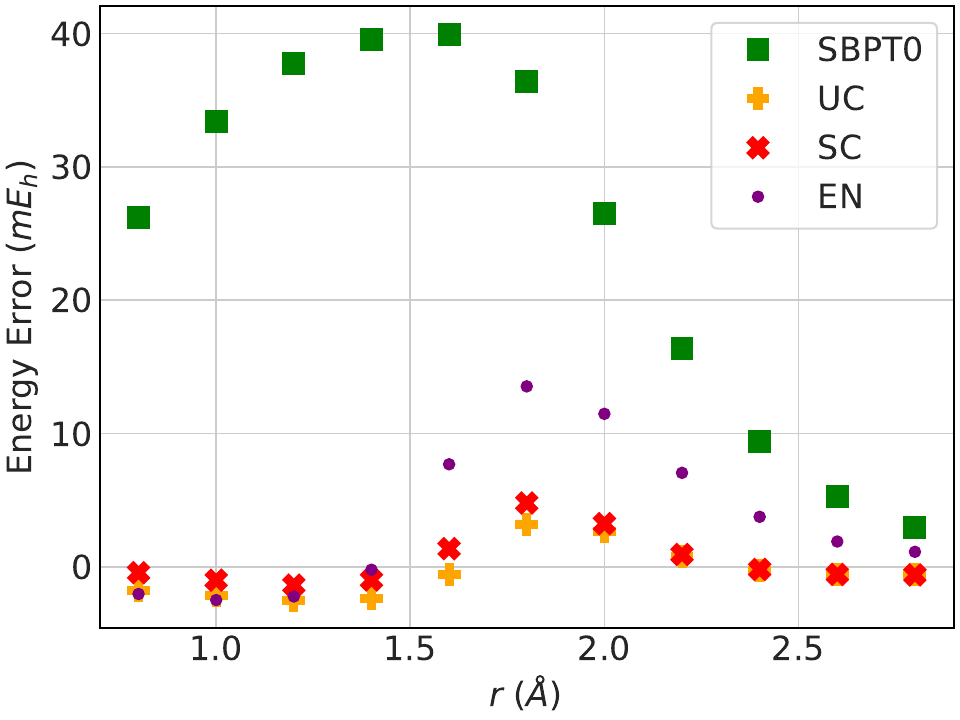}
    \caption{SBPT2}
    \label{fig:n2_sbpt}
  \end{subfigure}
  \vfill
  \centering
  \begin{subfigure}[b]{0.48\textwidth}
    \centering
    \includegraphics[width=\textwidth]{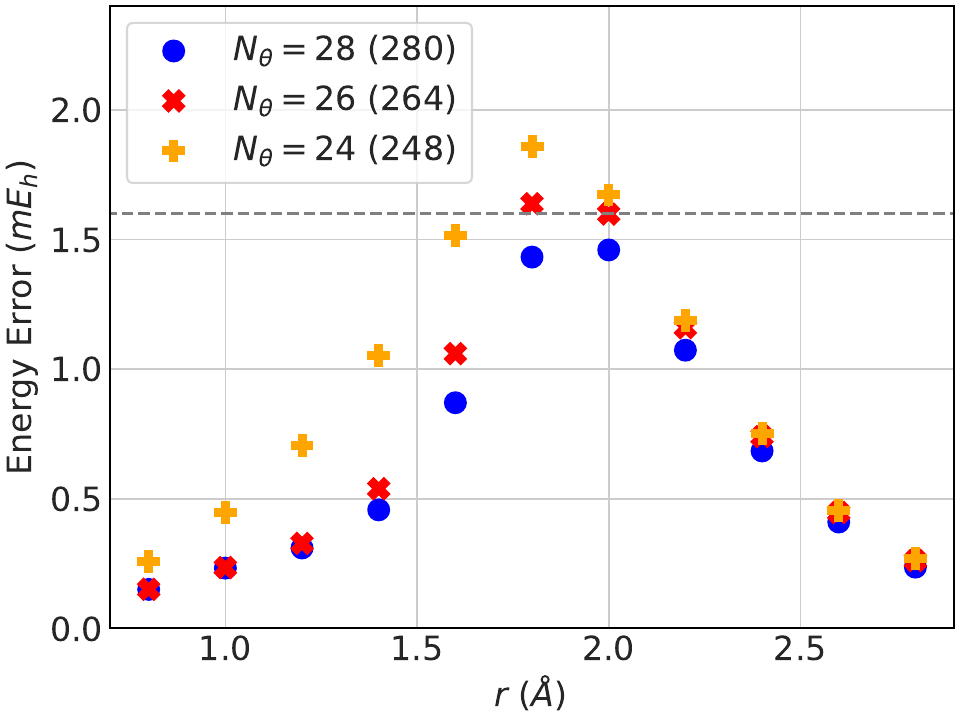}
    \caption{SCI-SBPT2: $\epsilon_2 = 0$}
    \label{fig:n2_sci1}
  \end{subfigure}
  \hfill
  \centering
  \begin{subfigure}[b]{0.48\textwidth}
    \centering
    \includegraphics[width=\textwidth]{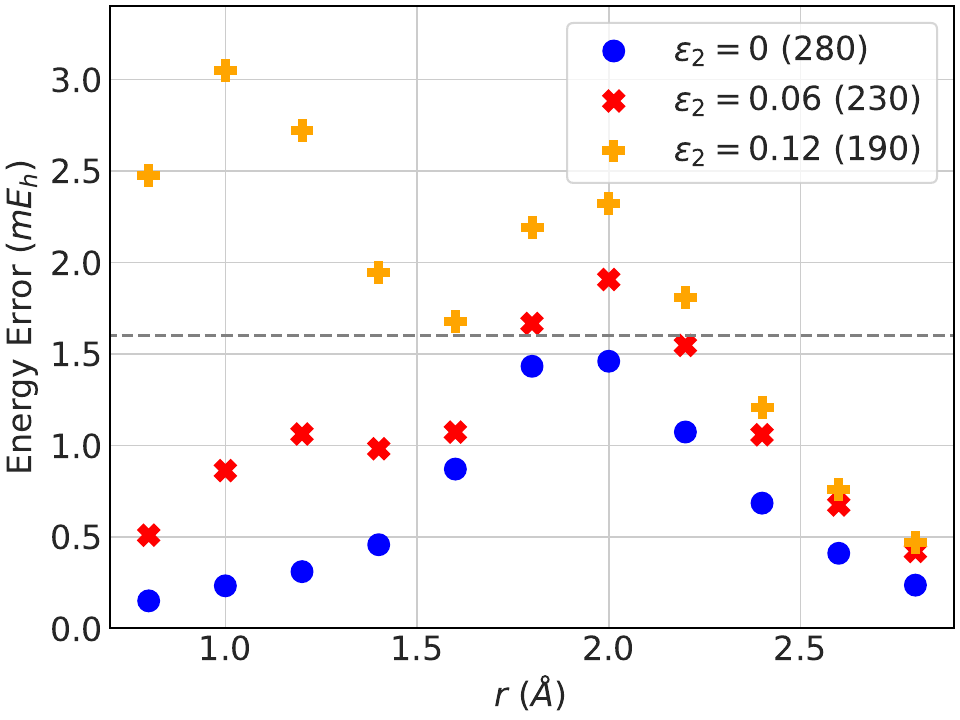}
    \caption{SCI-SBPT2: $\epsilon_1 = 0$}
    \label{fig:n2_sci2}
  \end{subfigure}
  \caption{The ground state energy dissociation curve of
    $\mathrm{N_2}$ in the STO-3G basis with various approximations
    similar to Fig.~\ref{fig:h2o_energies}.}
  \label{fig:n2_energies}
\end{figure}

Next we compute the ground state energy dissociation curve for the
nitrogen molecule in the STO-3G basis.  It is well known that the
dissociation of the molecule gives rise to multireference electronic
wavefunctions, and therefore it is an ideal problem to compare
different MRPTs. In the STO-3G basis with two core orbitals frozen,
there are eight molecular orbitals and ten electrons. The dissociation
curve of the ground state energy is plotted in
Fig.~\ref{fig:n2_dissociation}. We treat this as the exact solution,
and we compare it with various perturbative approximations at 11
values of $r$ equally spaced from $r = 0.8$ to $2.8$. Figure
\ref{fig:n2_dissociation} again shows that the SRPTs and MRPTs with a
small active space break down at large bond lengths.

The point group symmetry of the $\mathrm{N_2}$ molecule is
$D_{\infty h}$. There are three $\mathbb{Z}_2$ subgroups associated
with the reflection symmetries $\sigma_{xy}$, $\sigma_{xz}$, and
$\sigma_{yz}$. We group the orbitals that transform nontrivially under
these reflections as shown in Fig.~\ref{fig:n2} for $r=1.8$ and label
the grouping as $A_1(\sigma_{xy})$, etc. The lowest orbital in the
core belongs to $A_1(\sigma_{xy})$, but we omit it in the figure to
avoid cluttering. In the exact case, the number of configurations that
respect these three $\mathbb{Z}_2$ symmetries, as well as the fixed
number of spin up and down electrons, is $396$. Since there are eight
orbitals, the Jordan-Wigner transformation gives rise to $16$
qubits. Using the five $\mathbb{Z}_2$ symmetries, we can reduce the
number of required qubits to $11$ using the qubit tapering. The
summary of the resource requirements is given in Table \ref{table:n2}.

\begin{table}[h!]
  \centering
  \begin{tabular}{| c || c c c c |}
    \hline
    & Exact & CAS(6,6) & CAS(6,5) & SBPT \\ [0.5ex]
    \hline\hline
    Number of orbitals ($M$) & 8 & 6 & 5 & 6 \\
    \hline
    Number of qubits ($N_Q$) & 11 & 7 & 6 & 5 \\
    \hline
    Number of configurations ($N_{\rm{det}}$) & 396 & 56 & 16 & 32 \\ [1ex]
    \hline
  \end{tabular}
  \caption{The resource requirements---the number of qubits and
    configurations---for $\mathrm{N_2}$ in the STO-3G basis for the
    Hamiltonians defined in Fig.~\ref{fig:n2}. The exact case is the
    frozen core approximation, i.e., CAS(10,8). The CAS(6,5) is
    included for reference purposes.}
  \label{table:n2}
\end{table}

In the standard MRPT approaches, we can treat the two lowest orbitals
as external orbitals as they are weakly coupled. Using the six $2p$
orbitals shown in the middle figure of Fig.~\ref{fig:n2}, we construct
CAS(6,6) as the reference Hamiltonian. This gives a good agreement
with the exact solution as shown in Fig.~\ref{fig:n2_nevpt}, and in
fact it is the smallest CAS for NEVPT2 that gives reasonable results.
As we try to add additional external orbitals, they all give poor
results, since we need all $2p$ orbitals to capture the correct
physics.

We now apply SBPT2 to this problem.  As in NEVPT2, we treat the two
lowest orbitals as external orbitals.  As discussed in
Sec.~\ref{sec:approximate_symmetries}, we need to find orbitals that
are nontrivially coupled. Similar to the case of $\mathrm{H_2O}$, we
can take the groupings, $A_1(\sigma_{yz})$ and $A_1(\sigma_{xz})$, and
split each grouping into half as shown in the right figure of
Fig.~\ref{fig:n2}, using the same norm and chemistry arguments as
before. As a result, there are two additional exact $\mathbb{Z}_2$
symmetries in the reference Hamiltonian, which is equivalent to
reducing one additional orbital out of CAS(6,6) in terms of the qubit
count. Using SBPT, we see that the second-order correction gives good
agreement with the exact solution as shown in Fig.~\ref{fig:n2_sbpt}.
For both SBPT2 and NEVPT2, the mean-field treatment of the excited
states, the EN approximation, gives larger errors and deviations from
the full UC second-order correction compared to the SC approximation.
This suggests that $\mathrm{N_2}$ has a stronger multireference
character than $\mathrm{H_2O}$. The plot shows again that SBPT2 can
perform as good as NEVPT2 but with smaller resource requirements for
configurations and qubits.

Figures \ref{fig:n2_sci1} and \ref{fig:n2_sci2} show the results of
SCI-SBPT for various cutoffs in Eqs.~\eqref{epsilon_1} and
\eqref{epsilon_2}, where the dotted line indicates the kcal/mol
accuracy, $1.6 mE_h$. Again we select configurations based on one
value of $r$ and use the same configurations for the other values of
$r$ for simplicity. We have chosen $r=1.8$ as the selection point
because this is where the system exhibits multireference character.
With zero cutoffs $\epsilon_1 = \epsilon_2 = 0$, there are 280
configurations in 28 irreps out of 396 configurations in 55
irreps. Both figures show a consistent improvement as we decrease the
cutoffs, and the error is always positive, demonstrating that the
method is both convergent and variational. From
Fig.~\ref{fig:n2_sci1}, we see that all 28 irreps are important to
achieve the accuracy, suggesting that CAS(6,6) is indeed the minimum
active space to capture the correct physics. Within the 28 irreps, we
can still select important configurations using the $\epsilon_2$
cutoff, reducing the number of configurations from 280 to around 230
to achieve the accuracy for $r=1.8$, as shown in
Fig.~\ref{fig:n2_sci2}. The exact diagonalization of the ground state
energy shows that we need at most 140 out of 396 configurations to
achieve the accuracy for $r=1.8$.  This suggests that the selection of
28 irreps by SCI-SBPT2 misses some important configurations and
requires us to select more configurations to achieve the same
accuracy. This is another indication that the problem has a strong
multireference character. Also the plots show that the selected
configurations at one point may not work well at different points.  We
could overcome this problem by selecting important configurations from
different values of $r$.

\begin{figure}
  \centering
  \begin{subfigure}{0.48\linewidth}
    \centering
    \includegraphics[width=\linewidth]{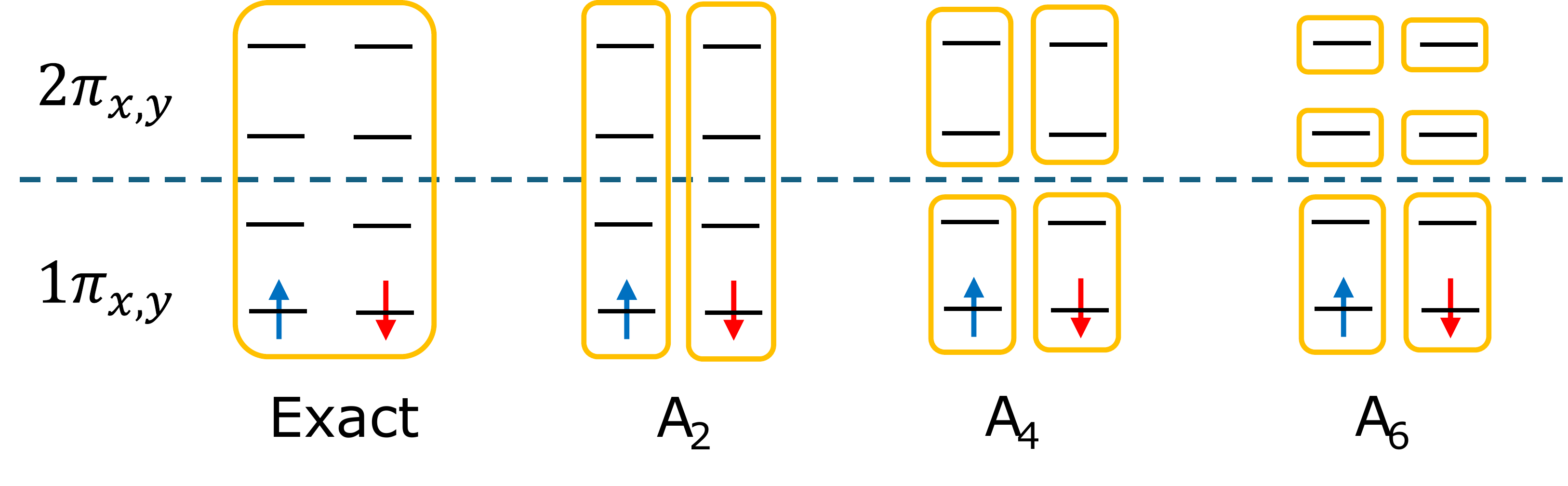}
    \caption{Groupings based on $\sigma_{xz}$ and $\sigma_{yz}$.}
    \label{fig:n2_631g_grouping}
  \end{subfigure}
  \hfill
  \centering
  \begin{subfigure}{0.48\linewidth}
    \centering
    \includegraphics[width=\linewidth]{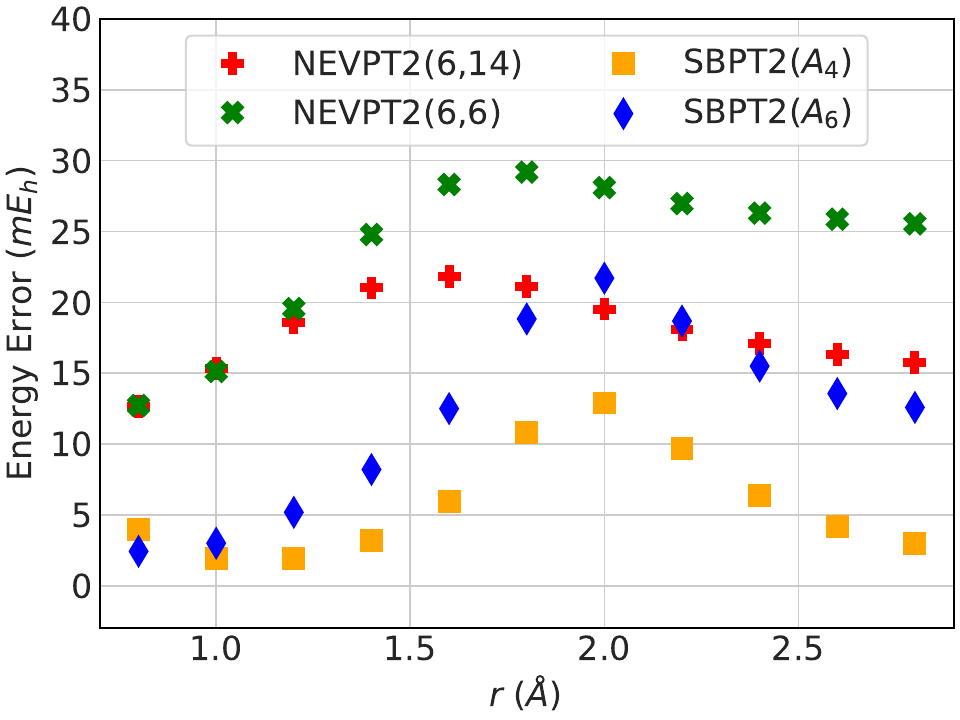}
    \caption{NEVPT2 and SBPT2}
    \label{fig:n2_631g_mrpt}
  \end{subfigure}
  \caption{$\mathrm{N_2}$ in the 6-31G basis. (a) Schematic orbital
    energy diagrams of the $\pi$ orbitals with various groupings used
    in SBPT. (b) NEVPT2 and SBPT2 results for the ground-state energy
    dissociation curve. We have used PySCF for NEVPT2 and the strongly
    contracted approximation in Eq.~\eqref{sc} with two different
    groupings, $A_4$ and $A_6$, for SBPT2.}
  \label{fig:n2_631g}
\end{figure}

Lastly, we show some results of the ground state energy dissociation
curve of $\mathrm{N_2}$ in the 6-31G basis. In this case, there are 18
orbitals and 14 electrons. By freezing two core orbitals, we have 16
orbitals and 10 electrons. We treat the freezing core approximation as
the exact solution and compare it with NEVPT2 and SBPT2
approximations. The resource requirements are shown in Table
\ref{table:n2_631g}. From Fig.~\ref{fig:n2_631g_mrpt}, we find that
although NEVPT2(6,6) gives a reasonable result with the same
computational resources as before, it struggles in the dissociation
limit. The larger active space size NEVPT2(6,14), which includes $3s$
and $3p$ orbitals, does not give a significant improvement.  We have
also found that NEVPT2(6,12) gives a similar result as
NEVPT2(6,14). This suggests that all 16 orbitals need to be included
in order to have a better result than NEVPT2.

\begin{table}[h!]
  \centering
  \begin{tabular}{| c || c c c c c |}
    \hline
    & Exact & CAS(6,14) & SBPT($A_4$) & SBPT($A_6$) & CAS(6,6)\\
    [0.5ex]
    \hline\hline
    Number of orbitals ($M$) & 16 & 14 & 16 & 16 & 6 \\
    \hline
    Number of qubits ($N_Q$) & 27 & 23 & 21 & 17 & 7 \\
    \hline
    Number of configurations ($N_{\rm{det}}$) & 2,388,528 & 16,584 & 41,472
                                      & 25,088 & 56 \\ [1ex]
    \hline
  \end{tabular}
  \caption{The resource requirements---the number of qubits and
    configurations---for $\mathrm{N_2}$ in the 6-31G basis. SBPT2 uses
    the groupings, $A_4$ and $A_6$, shown in
    Fig.~\ref{fig:n2_631g_grouping}. The exact case is the frozen core
    approximation, i.e., CAS(10,16).}
  \label{table:n2_631g}
\end{table}

We therefore perform SBPT2 on CAS(10,16). As mentioned above, there
are three $\mathbb{Z}_2$ subgroups associated with the three
reflection symmetries $\sigma_{xy}$, $\sigma_{xz}$, and
$\sigma_{yz}$. The orbitals that transform nontrivially under these
symmetries are doubled due to the $3s$ and $3p$ orbitals. For
simplicity, we consider $\sigma_{xz}$ and $\sigma_{yz}$ symmetries,
where the $\pi_{x,y}$ orbitals, which are linear combinations of
$2p_{x,y}$ and $3p_{x,y}$, transform nontrivially, as shown in
Fig.~\ref{fig:n2_631g_grouping}. The study of the STO-3G basis would
suggest that we could augment the symmetry using the grouping
$A_2$. In this case, we only have two extra symmetries as in the
STO-3G basis, while the number of orbitals is doubled. This would not
be favorable for scaling.

In this larger basis set, it is clear that we now have more ways to
group the orbitals. The grouping $A_4$ in
Fig.~\ref{fig:n2_631g_grouping} is one of them. It is natural to group
the $1\pi_{x,y}$ orbitals as in $A_4$ using the argument from the
STO-3G case. As for $2\pi_{x,y}$, we have chosen this grouping because
it gives the smaller norm of the perturbation than other groupings,
such as $A_2$ in Fig.~\ref{fig:grouping}. We note, however, that there
are two $\pi$ orbitals, unlike in the previous examples, and it is not
always clear how to separate $1\pi$ and $2\pi$ as they are the linear
combinations of the $2p$ and $3p$ orbitals.  This gives rise to size
inconsistency in the dissociation limit for the grouping $A_4$, unless
we enforce a clear separation between $1\pi$ and $2\pi$. The result is
plotted as the orange square in Fig.~\ref{fig:n2_631g_mrpt}, which
shows a better result than the NEVPT2 results. In this case, we have
six additional symmetries, compared to two in the STO-3G basis. Even
though the number of orbitals in CAS(10,16) is nearly triple that of
CAS(6,6) in the STO-3G basis, the number of additional symmetries is
also triple. As we further increase the number of symmetries as in
$A_6$, the result gets worse, showing that the contribution of
$2\pi_{x,y}$ orbitals is important for better accuracy.  This example
demonstrates that the number of symmetries can increase in SBPT2 as we
expand the basis set.

\section{Conclusions}
\label{sec:conclusions}

We have developed a multireference perturbation theory based on
symmetries of the electronic Hamiltonian.  This symmetry-based
perturbation theory (SBPT) extends existing perturbation theories to
include other forms of the reference Hamiltonian.

We have argued that we can use SBPT when some spin orbitals are weakly
coupled due to approximate symmetries.  There is an optimal way to
construct the reference Hamiltonian based on the symmetries, and in
this construction we can use the strongly contracted method, which
gives a scalable approximation to higher-order corrections.

We have formulated SBPT in the context of qubit operators for quantum
computing applications. The augmented $\mathbb{Z}_2$ symmetry can
reduce the qubit count via qubit tapering. We have also applied SBPT2
to SCI in order to overcome the shortcomings of perturbation theories
and obtained the convergent and variational results.

We have applied our methods to $\mathrm{N_2}$ and $\mathrm{H_2O}$ in
the STO-3G basis. By solving these small systems, we have explained
our approach in detail and shown that SBPT2 can give accurate results
with fewer configurations and qubits than NEVPT2.  We have further
studied $\mathrm{N_2}$ in the 6-31G basis and found that we can add
more symmetries in the reference Hamiltonian in the larger basis set,
providing an evidence that SBPT2 can be effective for larger
systems. The systematic exploration of how SBPT2 results depend on the
construction of the reference Hamiltonian in larger molecular systems
needs further investigation.

\section*{Acknowledgements}

We would like to thank Jay Lowell and Marna Kagele in the DC\&N
organization at Boeing for supporting this work.  We acknowledge Julia
Rice, Ben Link, and Triet Friedhoff for numerous helpful discussions
and James Shee and Francesco Evangelista for valuable feedback on the
manuscript.

\bibliographystyle{unsrt}
\bibliography{sbpt}

\end{document}